\documentclass[twocolumn,showpacs,preprintnumbers,amsmath,amssymb,floatfix,nofootbib]{revtex4}
\usepackage{graphicx}
\usepackage{dcolumn}
\usepackage{bm}
\usepackage{color}
\usepackage{url}
\usepackage{amsmath}
\usepackage{subfigure}
\usepackage{cleveref}

\graphicspath{{./figures/}}
\newcommand{\be}{\begin{equation}}
\newcommand{\ee}{\end{equation}}
\newcommand{\ba}{\begin{eqnarray}}
\newcommand{\ea}{\end{eqnarray}}
\newcommand{\Mc}{{\cal M}}
\newcommand{\Ms}{M_{\odot}}

\newcommand{\T}{\vec{\theta}}
\newcommand{\event}{\epsilon}
\newcommand{\hyp}{\mathcal{H}}
\newcommand{\cosmoP}{\vec{\Omega}}
\newcommand{\di}{\mathrm{d}}
\newcommand{\info}{\mathcal{I}}
\def\ltsima{$\; \buildrel < \over \sim \;$}
\def\simlt{\lower.5ex\hbox{\ltsima}}
\def\gtsima{$\; \buildrel > \over \sim \;$}
\def\simgt{\lower.5ex\hbox{\gtsima}}
\def\ksM{km s$^{-1}$ Mpc$^{-1}$}
\begin{document}

\title[Inference of the cosmological parameters with GW]{Inference of the cosmological parameters from gravitational waves: application to second generation interferometers}

\author{Walter \surname{Del Pozzo}$^{1,2}$}

\affiliation{$^{1}$Nikhef, National Institute for Subatomic Physics, Science Park 105, 1098 XG Amsterdam, The Netherlands}
\email{walterdp@nikhef.nl}
\affiliation{$^{2}$School of Physics and Astronomy, University of Birmingham, Edgbaston, Birmingham B15 2TT, UK}

\date{today}
\begin{abstract}
The advanced world-wide network of gravitational waves (GW) observatories is scheduled to begin operations within the current decade. Thanks to their improved sensitivity, they promise to yield a number of detections and thus to open a new observational windows for astronomy and astrophysics. Among the scientific goals that should be achieved, there is the independent measurement of the value of the cosmological parameters, hence an independent test of the current cosmological paradigm. Due to the importance of such task, a number of studies have evaluated the capabilities of GW telescopes in this respect. However, since GW do not yield information about the source redshift, different groups have made different assumptions regarding the means through which the GW redshift can be obtained. These different assumptions imply also different methodologies to solve this inference problem. This work presents a formalism based on Bayesian inference developed to facilitate the inclusion of all assumptions and prior information about a GW source within a single data analysis framework. This approach guarantees the minimisation of information loss and the possibility of including naturally event-specific knowledge (such as the sky position for a Gamma Ray Burst - GW coincident observation) in the analysis. 
The workings of the method are applied to a specific example, loosely designed along the lines of the method proposed by Schutz in 1986, in which one uses information from wide-field galaxy surveys as prior information for the location of a GW source. I show that combining the results from few tens of observations from a network of advanced interferometers will constrain the Hubble constant $H_0$ to an accuracy of $\sim 4 - 5$\% at 95\% confidence. 

\end{abstract}

\pacs{95.85.Sz,98.80.-k,04.30.-w}

\maketitle

\section{Introduction} 
 
The current decade will see the beginning of the era of gravitational waves astronomy. A world-wide network of second-generation interferometric gravitational waves (GW) detectors is in fact scheduled to begin operations in 2014--2015. Currently, the already existing LIGO facilities in USA \citep{advligo} and Virgo in Italy \citep{advvirgo} are in the process of being upgraded and the Large Cryogenic Gravitational-wave Telescope (LCGT) \citep{lcgt}  in Japan and possibly IndiGO \cite{indigo} in India are supposed to join the global network in following years. 
Thanks to their increased sensitivity, second-generation instruments are expected to yield several positive detections of compact binary systems coalescences; the detection rate is estimated to be in the range $\sim 1-100$ yr$^{-1}$, depending on the actual astrophysical event rate, instrument duty cycles and sensitivity evolution \citep{cbc-low-mass-S5VSR1}.

Among the many possibilities offered by a new observational window, gravitational waves (GW) from coalescing compact binaries potentially offers a one-step-only, totally independent measurement of the Hubble (and other cosmological) parameters, as pointed out by \citet{Schutz:1986} over 25 years ago. Differently from electro-magnetic observations, where one has to resort to cross-calibration of multiple distance indicators, for GW observations the luminosity distance is a \emph{direct observable} \citep{Schutz:1986,ChernoffFinn:1993,SathyaSchutz:2009}, and if one could infer from other means the redshift of the source, one could estimate the cosmological parameters from the luminosity distance--redshift relation. As second-generation (or advanced) ground-based gravitational-wave laser interferometers are being installed, this becomes a very concrete scenario, which may contribute to the solution of yet unresolved issues both in the determination of the Hubble constant \citep[see][for a review]{Jackson:2007} and in our understanding of the high redshift universe, and its mass-energy content. Several studies have already proven that space based observatories, such as the Laser Interferometric Space Antenna (LISA)\cite{lisa}, can successfully address both these issues: by measuring the redshift statistically, in \citet{MacLeodHogan:2008} it has been shown that $H_0$ can be determined with percent accuracy from the observation of extreme mass ratio systems and, using a similar approach, in \citet{PetiteauEtAl:2011} it has been shown that also $w$, the Dark Energy equation of state, can be accurately measured once the remaining cosmological parameters are known. Their analysis was greatly facilitated by the very good sky localisation capabilities offered by LISA for extreme mass ratio and massive binary black holes systems. 
 
The possibilities that ground based observatories offer have also been extensively investigated. In particular it has been shown that designed third generation interferometers will offer measurements of the Dark Energy equation of state that are competitive with current electro-magnetic measurements \cite{SathyaEtAl:2010,ZhaoEtAl:2011}, while second generation interferometers could plausibly constrain $H_0$  \citep{NissankeEtAl:2010,TaylorEtAl:2011}. 
The challenge in ground based GW observations is to obtain a redshift measurement for a GW detected source. As the GW error-box is $\approx 1-100$ deg$^2$, direct redshift measurements may be very challenging, despite optimistic assumptions made by several authors. For example, if short-lived Gamma Ray Bursts (GRB) are associated to the merger of compact objects with at least a neutron star component, this would provide such a measurement \citep{NissankeEtAl:2010,SathyaEtAl:2010,ZhaoEtAl:2011}. However, regardless of the still open debate on whether short GRB progenitors are indeed compact binaries, the fraction of coalescing systems producing a short GRBs might be as low as $10^{-2} $  \citep{BelczynskiEtAl:2008}. These considerations bear the question of whether it is at all feasible to use GW binaries as a new class of standard candles if the electro-magnetic counterpart it is not known. Recent literature focused on this possibility. \citet{TaylorEtAl:2011} explored the case in which the mass function of neutron stars is known, while \citet{MessengerRead:2011} suggest a direct measurement of the redshift from GWs, but the required sensitivity is achievable only by third generation instruments. Both methods aim at measuring directly the rest masses of the system and, by comparing with the observed --redshifted-- mass, extract the redshift of the source. 
Each of the aforementioned methods has its merits and its shortcomings: the difficulties of using GRBs to obtain the redshift have already been mentioned while the two latter approaches rely on the knowledge either of the intrinsic mass function of neutron stars or on the knowledge of their equation of state. Even if you would somehow acquire this information (possibly from GW, for example \cite{PannaraleEtAl:2011}), the inference of the cosmological parameters would be possible only for systems in which at least one of the two components is a neutron star. The implications are twofold: (i) we would be able to use less systems than observed, thus we would have less systems to average out systematics that could affect our estimates; (ii) our distance reach, and thus our sensitivity to the energy density parameters, would be seriously reduced. 

In view of these problems, this paper will present a formalism based on Bayesian inference aimed at measuring the cosmological parameters using GW for any particular cosmological model under consideration. This general formalism allows to take into account all information that is available and relevant for all GW detections and as such it is widely applicable to any particular ``class'' of putative standard sirens. To exemplify the workings of the method, this study will investigate the capabilities of the upcoming network of advanced ground-based observatories in a similar scenario as the one proposed originally by Schutz \cite{Schutz:1986} and further developed by MacLeod \& Hogan \cite{MacLeodHogan:2008} for LISA. In particular, we will see that a three advanced interferometers network will be able to constrain $H_0$ independently from any electro-magnetic measurement as $0.679\leq h \leq 0.722$ with as few as 20 events and as $0.686 \leq h \leq 0.714$ with 50.  This kind of accuracies are comparable with results from the Hubble Key Project \cite{FreedmanEtAl:2001} and from 7 years WMAP \cite{KomatsuEtAl:2011}.
The rest of the paper is organised as follows: Section \ref{s:method} presents the method in its generality and then proceeds in specialising it to the case of compact binary coalescences observed from second generation interferometers in conjunction with wide-field sky surveys. Section \ref{s:catalog} presents the GW catalogue on which the simulation in Section \ref{s:results} is based. Finally, the results are summarised and discussed in Section \ref{s:conclusions}. 

\section{Method}\label{s:method} 

This section is divided into two subsections. In the first one, I will introduce the formalism in its generality. In the second subsection, I will present an adapted version of the full formalism to estimate the cosmological parameters \emph{a posteriori}, that is after a measurement of the parameters describing a GW model. For the computation of the response to a GW of a network of detectors, I will use the same geometric conventions introduced in \cite{CutlerFlanagan:1994,AndersonEtAl:2001,NissankeEtAl:2010}. For a thorough discussion about these topics, the reader is referred to \cite{VitaleZanolin:2011} and references therein. The details of the detector networks that will be studied can be found in \cite{NissankeEtAl:2010,VitaleZanolin:2011,Schutz:2011}.

\subsection{Inference of the cosmological parameters from gravitational waves: the general approach}\label{ss:gen}

Consider a catalogue of gravitational wave events $\mathcal{E}\equiv\event_1,\ldots,\event_n$ observed by a network of $K$ gravitational wave detectors. The posterior probability distribution for the cosmological parameters -- the Hubble constant $H_0$, the density parameters $\Omega_m$, $\Omega_k$, and $\Omega_\Lambda$, and so on -- that I collectively represent with the $\cosmoP$ -- given the ensemble of events $\mathcal{E}$ and a cosmological model (or hypothesis) $\hyp$ is: 
\begin{equation}\label{eq:posteriors-multi-events}
p(\cosmoP|\mathcal{E},\hyp,\info)  = p(\cosmoP|\hyp,\info) \frac{p(\mathcal{E}|\cosmoP,\hyp,\info)}{p(\mathcal{E}|\hyp,\info)}
\end{equation}
where $p(\cosmoP|\hyp,\info)$ is the prior probability distribution for $\cosmoP$ given the cosmological hypothesis $\hyp$ and $\info$ indicates all the background information that is relevant for the inference problem under consideration. If the gravitational wave events are considered independent, the likelihood $p(\mathcal{E}|\cosmoP,\hyp,\info)$ in Eq.~(\ref{eq:posteriors-multi-events}) can be rewritten as a product of the likelihoods for each single event $\event_i$:
\be\label{eq:single-event-likelihood}
p(\mathcal{E}|\cosmoP,\hyp,\info)  = \prod_{i=1}^{n} p(\event_i|\cosmoP,\hyp,\info)\,,
\ee
thus, Eq.~(\ref{eq:posteriors-multi-events}) reads:
\be
p(\cosmoP|\mathcal{E},\hyp,\info)  = p(\cosmoP|\hyp,\info) \prod_{i=1}^{n} \frac{p(\event_i|\cosmoP,\hyp,\info)}{p(\event_i|\hyp,\info)}\,.
\ee
Strictly speaking, the quantity $p(\event_i|\cosmoP,\hyp,\info)$ is not a real likelihood because it is the result of the marginalisation over the GW signal intrinsic parameters, that are \emph{nuisance} parameters for the purpose of inferring $\cosmoP$. Such quantity is sometimes referred to as \emph{quasi-likelihood} \cite{Jaynes}. If we indicate with $\T$ the set of parameters on which the GW waveform depends, the quasi-likelihood is:
\be\label{eq:quasi-likelihood}
p(\event_i|\cosmoP,\hyp,\info)=\int \di \T \, p(\T|\cosmoP,\hyp,\info)p(\event_i|\cosmoP,\T,\hyp,\info)\,,
\ee
in which I introduced the prior probability distribution $p(\T|\cosmoP,\hyp,\info)$ for $\T$ and the integral is to be performed over the full parameter space defined by $\T$. For non-spinning waveforms, in celestial coordinates, $\T \equiv (m_1,m_2,\phi_c,\tau_c,\alpha,\delta,\psi,\iota,z,D_L)$: the component masses $m_1$ and $m_2$, the -- geocentric -- phase $\phi_c$ and time $\tau_c$ at coalescence, right ascension $\alpha$, declination $\delta$, polarisation angle $\psi$, inclination angle $\iota$, the redshift $z$ and -- a priori -- the luminosity distance $D_L$. 
The prior probability distribution $p(\T|\cosmoP,\hyp,\info)$, thanks to the product rule, can be factorised into a product of the prior probabilities for each parameter or group of parameters:
\begin{widetext}
\ba\label{eq:joint-prior}
p(\T|\cosmoP,\hyp,\info)=p(m_1,m_2|\info)p(\phi_c|\info)p(\tau_c|\info)p(\psi,\iota|\info)p(z,\alpha,\delta|\info)p(D_L|z,\cosmoP,\hyp,\info)\,.
\ea
\end{widetext}
Since a cosmological model $\hyp$ predicts that $D_L$ is a function $D(\cosmoP,z)$ of the redshift $z$ and of the cosmological parameters $\cosmoP$, only two of the three parameters $D_L,z$ and $\cosmoP$ are independent. If we choose $\cosmoP$ and $z$ to be the independent parameters, what would be the prior for $D_L$ becomes: 
\be\label{eq:delta}
p(D_L|z,\cosmoP,\hyp,\info)=\delta(D_L-D(\cosmoP,z))\,,
\ee 
which shows explicitly that, when a cosmological model is considered, the luminosity distance $D_L$ is a not a model parameter.
The function $D(\cosmoP,z)$, in a Friedmann-Robertson-Walker-LeMa\^{i}tre universe, is given by \cite{Hogg:1999}:
\ba\label{eq:dl}
&&D(\cosmoP,z)=\nonumber\\
&&\left\{\begin{array}{ll}
	\frac{c(1+z)}{H_0}\frac{1}{\sqrt{\Omega_k}}\sinh[\sqrt{\Omega_k}\int_0^z\frac{dz^\prime}{E(z^\prime)}] & \mbox{for \, $\Omega_k > 0$} \\
	\frac{c(1+z)}{H_0}\int_0^z\frac{dz^\prime}{E(z^\prime)} & \mbox{for \, $\Omega_k = 0$} \\
	\frac{c(1+z)}{H_0}\frac{1}{\sqrt{|\Omega_k|}}\sin[\sqrt{|\Omega_k|}\int_0^z\frac{dz^\prime}{E(z^\prime)}] & \mbox{for \, $\Omega_k < 0$} \end{array} \right.
\ea 
and 
\be
E(z^\prime)=\sqrt{\Omega_m (1+z^\prime)^3+\Omega_k (1+z^\prime)^2+\Omega_\Lambda}\,.
\ee 

Simplifications to the problem of inferring $\cosmoP$ from GW are obtained by imposing restrictions to the form of some of the prior probability distributions in Eq.~(\ref{eq:joint-prior}). For instance, a coincident observation of a GRB and a GW \cite{HolzHughes:2005,NissankeEtAl:2010,SathyaEtAl:2010,ZhaoEtAl:2011} implies \emph{a priori} knowledge of the sky position and redshift of the GW source. Therefore, it is equivalent to imposing $p(z,\alpha,\delta|\info)=\delta(z-z_{GRB})\delta(\alpha-\alpha_{GRB})\delta(\delta-\delta_{GRB})$ where $\alpha_{GRB},\delta_{GRB}$ and $z_{GRB}$  are sky position and redshift of the observed GRB. Similarly, the use of the neutron star mass function to infer the redshift, as proposed recently by Taylor et al \cite{TaylorEtAl:2011}, corresponds to a very particular choice of $p(m_1,m_2|\info)$ that is the neutron star mass function. With the formalism presented herein, it is easy to see how to combine various assumptions and how it would be possible in the future to use a unique methodology for any kind of GW observation and optimise the relevant inference of $\cosmoP$ given any additional  information about the source at hand.

The prior distributions for the orientation angles $\iota$ and $\psi$ will be taken as uniform on the 2-sphere, and uniform on the time of coalescence $\tau_c$ in an interval of $\pm$ 1 second centred around the ``true'' coalescence time. The joint prior for redshift and sky position instead will be set by measurements extracted from wide-field sky surveys such as SDSS \cite{YorkEtAl:2000}. If we assume that a GW source is bound to be localised in a galaxy, then this assumption implies that the only admissible coordinates of a GW event are the ones corresponding to some galaxy. A wide-field survey can be seen as a list of redshifts and sky positions for all the galaxies that manage to be bright enough to be detected. This, in turn, defines a joint probability distribution of sky position and redshifts that we can take to be the joint prior required for Eq.~(\ref{eq:joint-prior}):
\be\label{eq:prior-z}
p(z,\alpha,\varphi|\info)\propto\sum_{j=1}^N p_j\delta(z-z_j)\delta(\alpha-\alpha_j)\delta(\delta-\delta_j)
\ee
where $N$ is the total number of galaxies identified by the survey and the $p_j$ are the weights one might assign to each galaxy. In what follows the weights will be taken to be equal for each galaxy, $p_j=1$ for $j=1,\ldots,N$. 

Let's turn our attention now to the likelihood $p(\event_i|\cosmoP,\T,\hyp,\info)$. If one assumes that the noise is independent and uncorrelated across different detectors, each of the single event likelihoods in Eq.~(\ref{eq:quasi-likelihood}) is further expressed as a product of the likelihoods for each of the detectors:
\be
p(\event_i|\cosmoP,\T,\hyp,\info) = \prod_{k=1}^{K} p(\event_i^{(k)}|\cosmoP,\T,\hyp,\info)\,.
\ee 
The likelihood at each detector is given by \cite{Finn:1992}:
\be\label{eq:likelihood}
p(\event_i^{(k)}|\cosmoP,\T,\hyp,\info) = e^{-(s_i^{(k)}-h^{(k)}(\cosmoP,\T)|s_i^{(k)}-h^{(k)}(\cosmoP,\T))/2}
\ee
in which the \emph{strain} $s_i^{(k)}$ for the $k$-th detector has been introduced and $(\ldots|\ldots)$ indicates the scalar product:
\be\label{e:scalar}
(\mathrm{f}|\mathrm{g})=2\int_0^{\infty}df\,\frac{\tilde{\mathrm{f}}^*\tilde{\mathrm{g}}+\tilde{\mathrm{f}}\tilde{\mathrm{g}}^*}{S_n^{(k)}(f)}
\ee
and $h^{(k)}(\cosmoP,\T)$ is the GW template. The $S_n^{(k)}(f)$ is the \emph{noise power spectral density} for the $k$-th detector. The signal-to-noise ratio $\rho_k$ in the $k$-th detector is defined in terms of the scalar product in Eq.~(\ref{e:scalar}) as $(h^{(k)}(\cosmoP,\T)|h^{(k)}(\cosmoP,\T))$. The network signal-to-noise ratio $\rho_{\mathrm{network}}$ is given by:
\be
\rho_{\mathrm{network}}=\sqrt{\sum_{k=1}^{K}\rho_k^2}\,.
\ee

\subsection{Inference of the cosmological parameters from gravitational waves: the \emph{a posteriori} approach}

The formalism presented in the previous section might be difficult to apply in practice, especially if, by the time of operation of the second-generation interferometers, there will be all-sky data without the necessary coverage and depth to be used for the definition of the prior in Eq.~(\ref{eq:prior-z}). However, the inference of $\cosmoP$ can still be done \emph{a posteriori}; after a 3-dimensional volume in the sky has been identified by means of standard data analysis pipelines, one would go and identify all possible hosts within the 3-dimensional volume and then proceed at inferring $\cosmoP$. In general, the pipelines currently used in the LIGO Algorithm Library produce a $n$-dimensional posterior probability distribution $p(\T^\prime|\event_i,\info^\prime)$ for all the parameters $\T^\prime$  -- that are the same as $\T$ but not including the redshift $z$, that is $\T\equiv\T^\prime \cup z$ -- describing the GW waveform. Note that the background information in this case is different from the previous subsection, $\info^{\prime}\neq\info$. From Bayes' theorem, the posterior for $\T^\prime$ is:
\be
p(\T^\prime|\event_i,\info^\prime)=p(\T^\prime|\info^\prime)\frac{p(\event_i|\T^\prime,\info^\prime)}{p(\event_i|\info^\prime)}\,.
\ee
When we want to infer $\cosmoP$ from $p(\T^\prime|\event_i,\info^{\prime})$, we suddenly ``remember'' that a cosmological model exists and that we want to measure the parameters $\cosmoP$ on which this model depends and that in our inference of $\T^\prime$ we have ignored. Formally, this logical process corresponds to the redefinition $\info^{\prime}=\cosmoP,\hyp,\info$. The joint posterior can then be rewritten as $p(\T^\prime|\event_i,\info^\prime)=p(\T^\prime|\event_i,\cosmoP,\hyp,\info)$. The quasi-likelihood in Eq.~(\ref{eq:quasi-likelihood}) becomes:
\be
p(\event_i|\cosmoP,\hyp,\info)=\int \di \T^\prime\di z \, p(\T^\prime,z|\cosmoP,\hyp,\info)p(\event_i|\cosmoP,\T^\prime,z,\hyp,\info)\,,
\ee
which can be further rewritten as:
\begin{widetext}
\be
p(\event_i|\cosmoP,\hyp,\info)=\int \di \T^\prime\di z \, p(z|\T^\prime,\cosmoP,\hyp,\info)p(\T^\prime|\cosmoP,\hyp,\info)p(\event_i|\cosmoP,\T^\prime,z,\hyp,\info)\,.
\ee
\end{widetext}
In the environment defined by $\info^\prime$, the knowledge of $z$ is irrelevant for $p(\event_i|\cosmoP,\T^\prime,z,\hyp,\info)$, therefore we can ignore the conditioning on $z$ and we are left with:
\ba
&&p(\event_i|\cosmoP,\hyp,\info)=\nonumber\\
&&\int \di \T^\prime\di z \, p(z|\T^\prime,\cosmoP,\hyp,\info)p(\T^\prime|\info^\prime)p(\event_i|\T^\prime,\info^\prime)\,,
\ea
in which we can recognise the ``standard'' likelihood $p(\event_i|\T^\prime,\info^\prime)$ and prior distribution $p(\T^\prime|\info^\prime)$. The term $p(z|\T^\prime,\cosmoP,\hyp,\info)$ is the prior distribution for $z$ once we assume $\T^\prime$ as known. The simplifying assumption that only $D_L,\alpha$ and $\delta$ are relevant to determine $z$ implies:
\be
p(z|\T^\prime,\cosmoP,\hyp,\info)\equiv p(z|D_L,\alpha,\delta,\cosmoP,\hyp,\info)
\ee
which corresponds to the selection of only the galaxies within the measured 3-dimensional volume in the sky, and thus coincides with the assumptions in \cite{Schutz:1986,MacLeodHogan:2008,PetiteauEtAl:2011}. We can then marginalise over all the remaining, and non-relevant, parameters and obtain
\begin{widetext}
\ba
&&p(\event_i|\cosmoP,\hyp,\info)=\nonumber\\
&&\int \di D_L \di\alpha\di\delta \di z \, p(z|D_L,\alpha,\delta,\cosmoP,\hyp,\info)p(D_L,\alpha,\delta|\cosmoP,\hyp,\info)p(\event_i|D_L,\alpha,\delta,\cosmoP,\hyp,\info)\,.
\ea
\end{widetext}
One can then proceed as in the previous subsection and impose the constraint on $D_L$ given by the luminosity distance -- redshift relation and finally calculate the posteriors for $\cosmoP$ as in Eq.~(\ref{eq:posteriors-multi-events}). 

\section{Gravitational wave events catalogue}\label{s:catalog}

\begin{figure}[h]
\centering
\includegraphics[angle=0,width=\columnwidth]{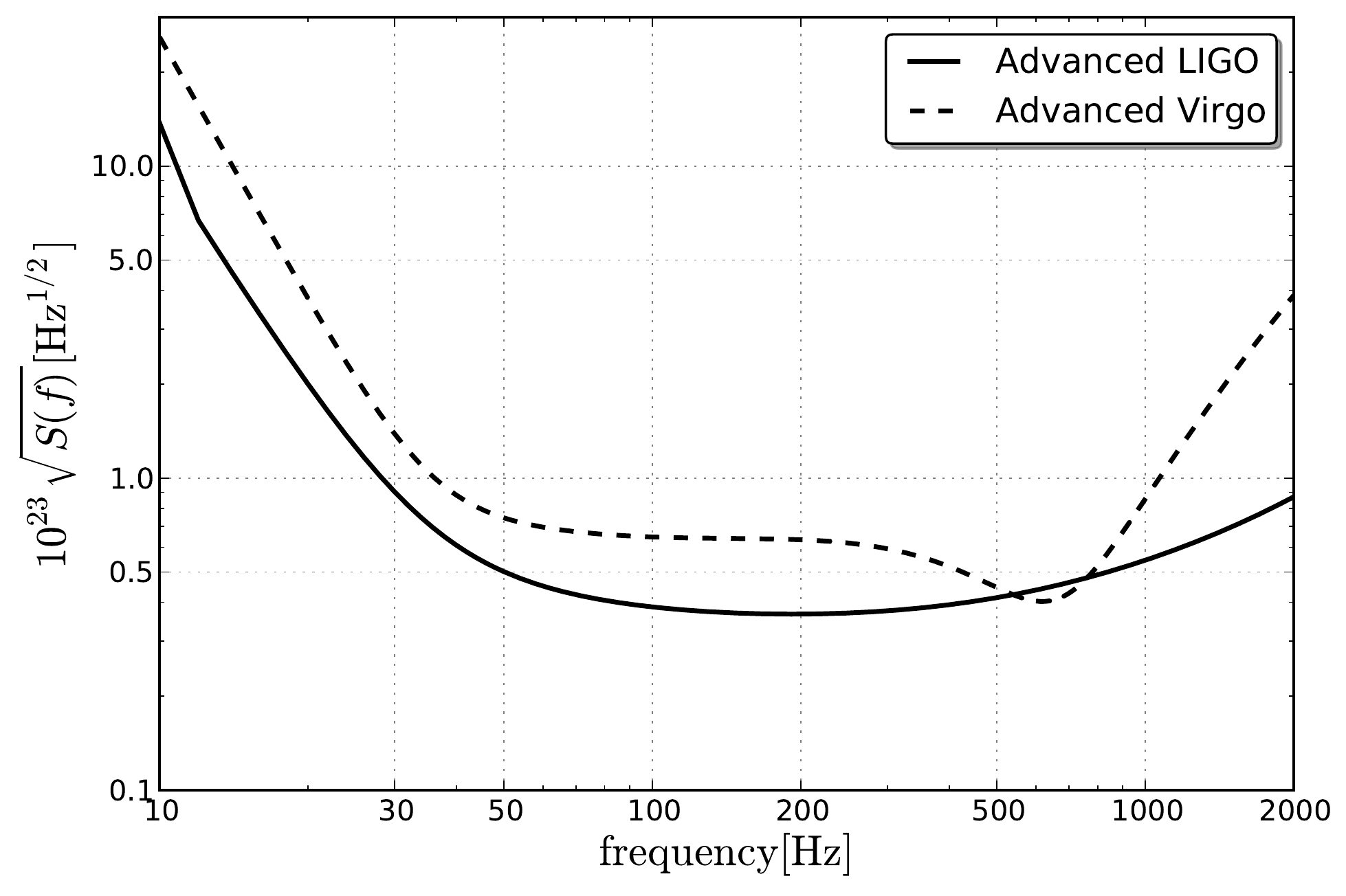}
\caption{The high-power, zero-detuning noise curve for Advanced LIGO, and the BNS-optimized Advanced Virgo noise curve.}
\label{fig:noisecurves}
\end{figure}

I applied the general formalism presented in section \ref{ss:gen} to a mock catalogue of GW events observed by a network of advanced interferometers in conjunction with the Sloan Digital Sky Survey Data Release 8\citep{SDSSDR8}. The SDSS DR8 spectroscopic catalogue comprises 840,375 galaxies and covers an area of 9,274 square degrees in the northern sky.
I considered three detector networks:
\begin{enumerate}
\item a three interferometers network made by LIGO Hanford, LIGO Livingston and VIRGO (HLV);
\item a four interferometers network made by LIGO Hanford, LIGO Livingston, VIRGO and LCGT (HLVJ);
\item a five interferometers network made by LIGO Hanford, LIGO Livingston, VIRGO in conjunction with LCGT and IndiGO (HLVIJ).
\end{enumerate} 
The locations and orientations of all the detectors considered can be found in \cite{NissankeEtAl:2010,Schutz:2011,VitaleZanolin:2011}. 
For the two Advanced LIGOs, for LCGT and IndiGO, I assumed the high-power, zero-detuning noise curve expected for Advanced LIGO, while for Advanced Virgo I assumed the BNS-optimized noise curve. These are given in Fig.~\ref{fig:noisecurves}.
For each event I simulated ten independent noise realisations and results will be shown as an average over them. When a GW is present, the strain in each detector, labeled by $k$, is given by the sum of the detector noise $n^{(k)}(f)$, assumed Gaussian with zero mean, and the GW $h^{(k)}(f)$:
\be
s^{(k)}(f) = n^{(k)}(f)+h^{(k)}(f).
\ee  
The response to the GW in the $k$-th detector is given by:
\ba
&&h^{(k)}(f)=\sum_{\mathrm{pol=+,\times}}D^{(k)}_{ij}e_{ij}(\mathbf{\hat{n}})h_{(\mathrm{pol})}(\T,\cosmoP;f) \\
&&\equiv e^{2\pi i\mathbf{r}_k\cdot \hat{\mathbf{n}}f} \left[F^{(k)}_+ h_+(\T,\cosmoP;f) + F^{(k)}_\times h_{\times}(\T,\cosmoP;f)\right]
\ea
where $D^{(k)}_{ij}$ is the \emph{detector tensor} and $\mathbf{\hat{n}}$ is the unit vector along which the GW propagates. In the second equality I introduced  the \emph{antenna pattern functions} $F_+^{(k)}$ and $F_\times^{(k)}$ for the $k$-th detector and the vector $\mathbf{r}_k$ that from the centre of the Earth points to the location of the $k$-th detector.
The GW signals are taken to belong to the TaylorF2 family \cite{BuonannoEtAl:2009}:
\be
h_{(\mathrm{pol})}(\T,\cosmoP;f)=A_{(\mathrm{pol})}(\T,\cosmoP)f^{-7/6}e^{i(2\pi \Delta\tau f + \Phi_{PN}(f) + \phi_c)}\,
\ee 
where $\Phi_{PN}(f)$ is the post-Newtonian (PN) expansion of the wave phase. For the purpose of this study, the PN phase has been restricted to the second order. 
As for the GW signal, the GW template used in the analysis is also taken to belong to the TaylorF2 family. The choice that the GW signal and the GW template belong to the same family implies that we are deliberately neglecting systematic effects due to signal -- template mismatch which might affect the end results.
The limits of integration for the computation of the likelihood, Eq.~(\ref{eq:likelihood}), are $f_{\mathrm{min}}=10\mathrm{Hz}$ and the frequency of the last stable circular orbit:
\be
f_{\mathrm{lso}}=\left(6^{3/2}\pi(m_1+m_2)\right)^{-1}
\ee
where $m_1$ and $m_2$ are the observed --redshifted-- component masses. 

The GW catalogue consists of a set of 1,000 compact binary coalescence events sampled from SDSS. Each galaxy is assigned equal probability of being the host of a GW event, therefore the redshift and sky position distributions of the GW events follow exactly the galaxy distributions. For each event the remaining parameters have been chosen as follows:
\begin{itemize}
\item the component masses of the binary system, in the system rest frame, are sampled from a uniform distribution $m_1,m_2 \in (1.0,15.0) M_\odot$;
\item the orientation angles are sampled from a uniform distribution on the 2-sphere;
\item the times of coalescence are evenly spaced by 1,000 seconds\footnote{This choice is motivated by the fact that this study does not take into account the rate, and consequently the distribution of arrival times of the GW signals. In principle, the rate, which also depends on the cosmology, could be simultaneously inferred and used as an additional constraint for the measurement of $\cosmoP$.}.
\end{itemize}
  
As SDSS covers about half of the northern sky, special care has to be taken when dealing with events that are close to the edge of the survey. For this reason, only GW events located at least $10$ degrees away from any of the edges were considered. Moreover, sources were restricted to have $z \leq 0.1$ ($\sim 460$ Mpc in a $\Lambda$CDM cosmology). This redshift corresponds to the $\sim$100\% completeness limit of SDSS for galaxies brighter than $17.77$ in the $r$ band, and therefore for the spectroscopic survey, and, given the most plausible rates \citep{cbc-low-mass-S5VSR1}, should yield 1--50 detections per year. Given the aforementioned redshift cut, the total number of galaxies actually considered in the analysis is 362,528. Because of the apparent magnitude cut for spectroscopic targets, data from SDSS beyond $z\simeq0.1$ must be used with caution. Since the detection efficiency drops by as much as $90$\% at $z\simgt0.15$, the galaxy number count seriously underestimates the actual number of host candidates. While for simulations, where one has complete control over the system, this might be a reasonable choice, in real observations the probability of missing the real host is unacceptably high. Thus the bias introduced in the estimate of $\cosmoP$ would be very significant with the result of drawing the wrong conclusions about the evolution of the Universe. A more realistic approach when assessing the performance of a GW observatory for high $z$ events is to use numerical N-body simulations to define the galactic population, as done in \cite{PetiteauEtAl:2011}, which do not suffer from incompleteness. 
%
%
\begin{figure}[ht]
\begin{tabular}{c}
\resizebox{\columnwidth}{!}{\includegraphics{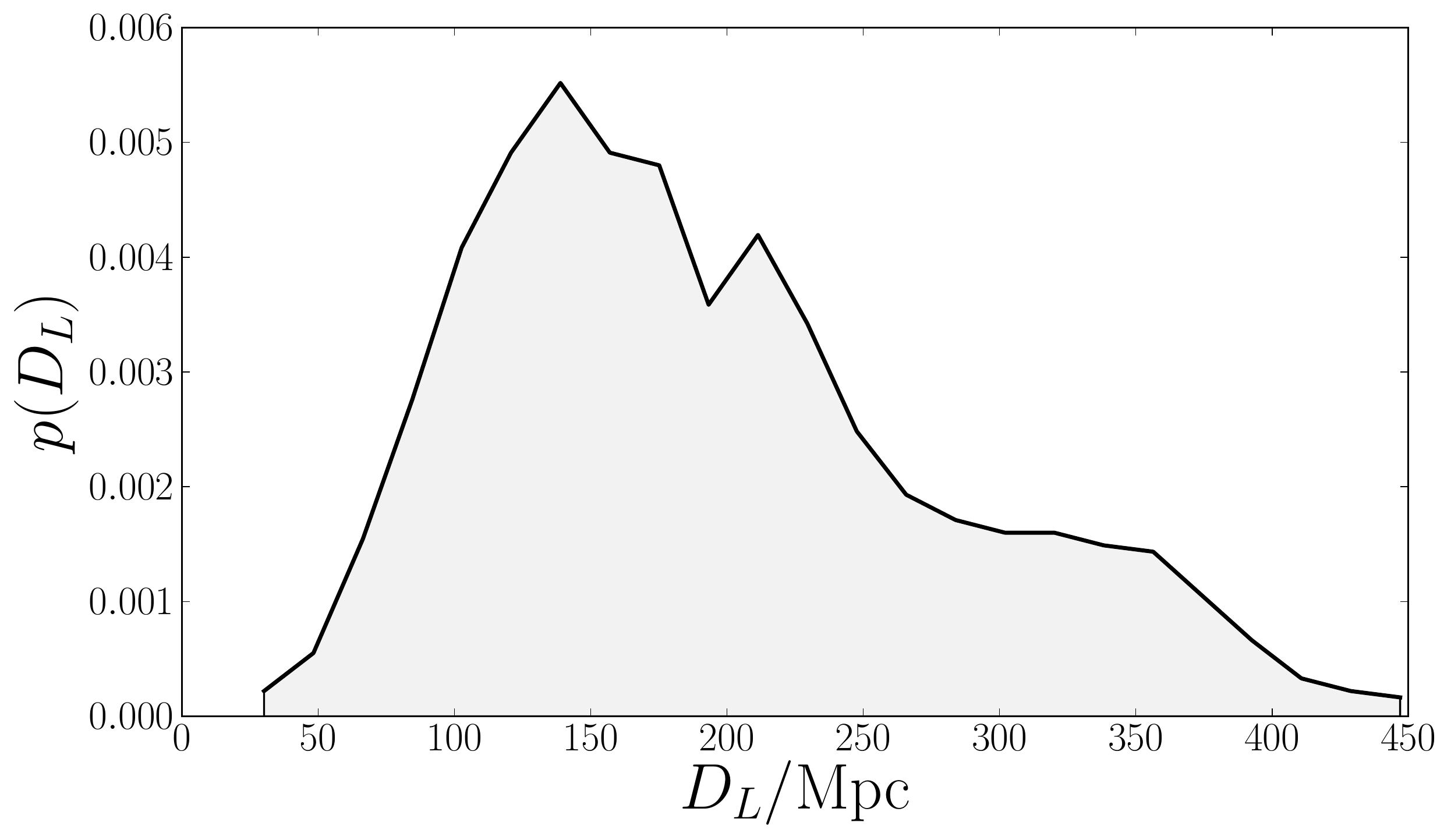}}\\
\resizebox{\columnwidth}{!}{\includegraphics{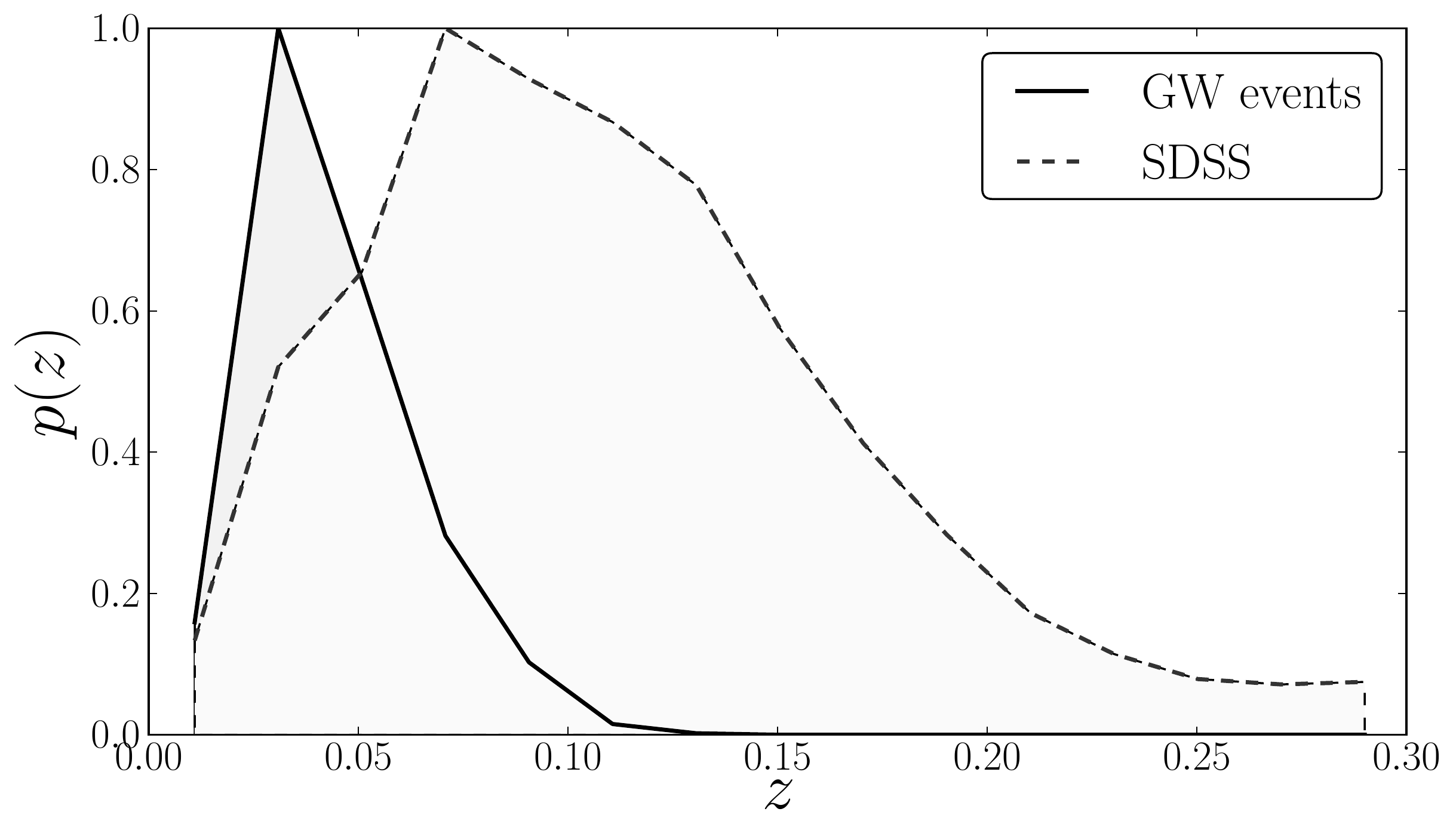}}\\
\resizebox{\columnwidth}{!}{\includegraphics{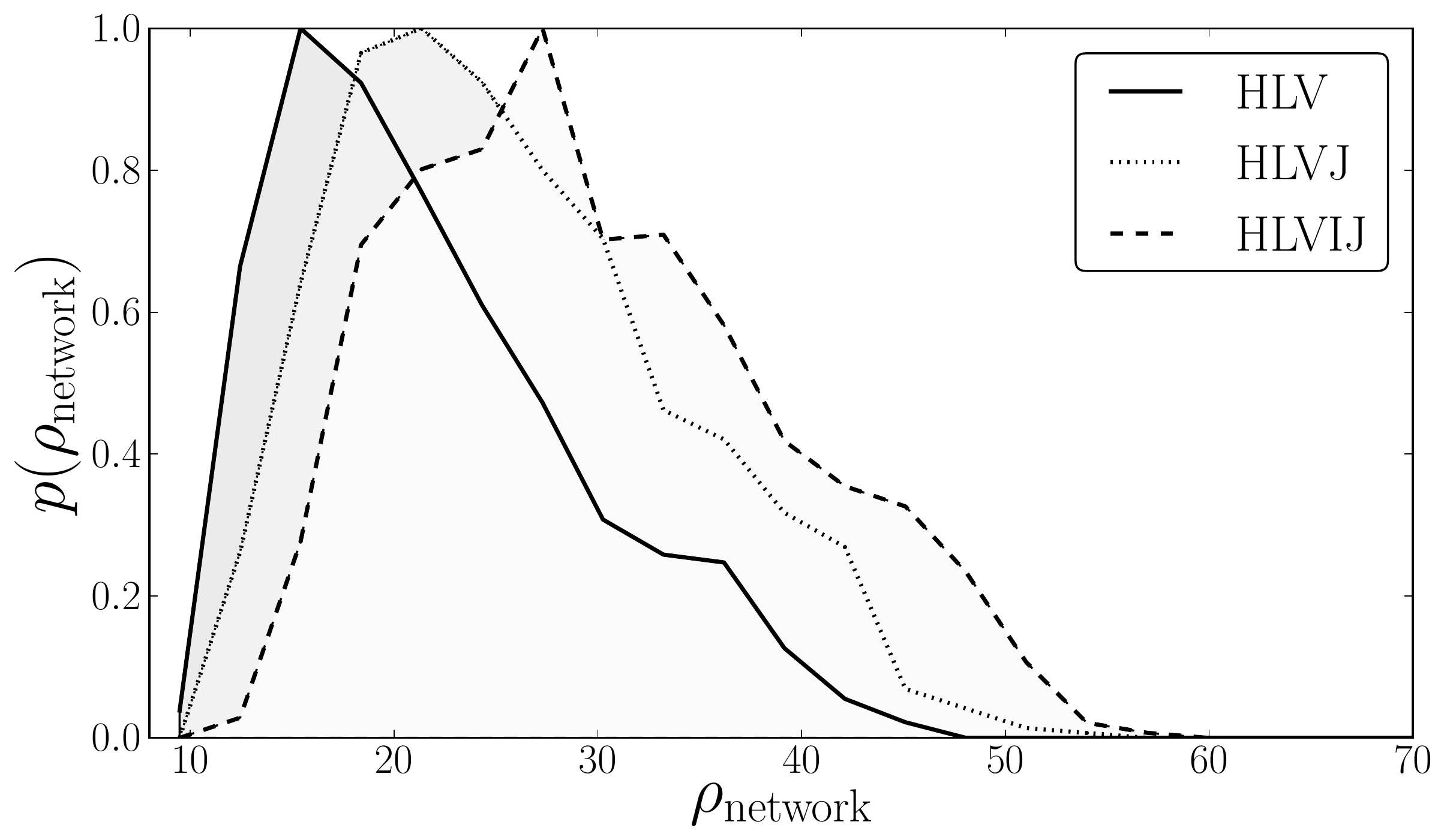}}
\end{tabular}
\caption{Properties of the 1,000 GW events catalogue. \emph{Top panel}: distribution of the distances $D_L$ calculated from a $\Lambda$CDM universe and the galaxy redshift. \emph{Centre panel}: distribution of the redshifts $z$ of the sources (solid line) compared to the overall SDSS redshift distribution (dashed line). \emph{Bottom panel}: distribution of the network signal-to-noise ratio $\rho_{\mathrm{network}}$ for the HLV network (solid line), the HLVJ network (dotted line) and the HLVIJ network (dashed line). The distributions for $z$ and for $\rho_{\mathrm{network}}$ have been scaled to facilitate the comparison.}\label{f:catalog}
\end{figure}
%
%
The sampled $z$ and $D_L$, which is calculated assuming a concordance $\Lambda$CDM cosmology with $\left\{h,\Omega_m,\Omega_k,\Omega_\Lambda\right\}=\left\{0.7,0.3,0.0,0.7\right\}$\footnote{As customary, I introduced $h\equiv H_0/100$ \ksM.}, are shown in Fig.\ref{f:catalog}, top and central panel, respectively.  The detection threshold was set at a signal-to-noise ratio of $5.5$ in each of the detectors part of the HLVIJ network. The distributions of network SNRs for the HLV, the HLVJ and HLVJI networks are also shown in Fig.\ref{f:catalog}, bottom panel. Please note that for the three different networks considered the GW events are the same. The mode of the redshift distribution for the GW catalogue is $\sim 0.03$, so \emph{a posteriori} we are justified in choosing SDSS as a baseline for the generation of the GW events and as a prior of $z,\alpha,\delta$. As already mentioned, SDSS is complete to $z\simlt0.1$, therefore all the events detected by second generation interferometers have very high probability of being hosted by a galaxy that is, or will be, identified by current, or near-future, wide-field surveys. A typical redshift $z\sim 0.03$ corresponds to a typical ``Hubble flow'' velocity of $\sim$ 9,000 km$\cdot$s$^{-1}$.  About 50\% of the whole galaxy population is found in bound associations such as clusters or groups, \emph{e.g.} \cite{DengEtAl:2008}. Typical peculiar velocities $v$, within these associations, are 100 km$\cdot$s$^{-1}\simlt v \simlt$1,000 km$\cdot$s$^{-1}$, so $\sim$1--10\% of the redshift is possibly not of cosmological origin but due to the proper motion of the host galaxy. Therefore, we can anticipate that if only very few events are observed, in addition to the bias possibly introduced by the well known inclination--distance degeneracy, the estimate of $\cosmoP$ and in particular of $h$ can be biased by a similar percentage. It is therefore of crucial importance the effective ``averaging'' obtained from the computation of the joint posterior distribution in Eq.~(\ref{eq:posteriors-multi-events}) from many events to minimise this potential source of bias. It is worth noting that, even if one does not aim at estimating $\cosmoP$, the proper motion of the host galaxy might affect similarly the estimate of the component masses, with potential pernicious effects on the reconstruction of mass dependent quantities. 

\section{analysis}\label{s:results} 

This section is divided into three subsections. The first one will present the set up of the data analysis simulation, in particular will reiterate over the prior probability distributions for the parameters of interest. The second subsection will present results for the case of a single GW event and will compare the performance of the different networks in sky localisation and in the estimation of $\cosmoP$. The third subsection will instead deal with the joint posterior distribution for $\cosmoP$ in the three networks.  

\subsection{Prior probabilities}

The analysis of each signal has been performed using a Nested Sampling algorithm \citep{Skilling:2004}. The parameters that are estimated for each signal are $\cosmoP \equiv {h,\Omega_m,\Omega_k,\Omega_\Lambda}$, with the boundary condition $\Omega_k+\Omega_m+\Omega_\Lambda=1$, the redshift $z$, the sky position $\alpha,\delta$, the orientation $\iota,\psi$, the chirp mass $\Mc$ and the symmetric mass ratio $\eta$ and the time of arrival at the geocentre $t_c$. The cosmological hypothesis $\hyp$ considered for the analysis is a Friedmann-Robertson-Walker-Lema\^{i}tre universe whose luminosity distance--redshift relation is given in Eq.~(\ref{eq:dl}). 
Some of the prior probabilities for all these parameters have already been briefly introduced in Section \ref{s:method}. For the sake of clarity, let's iterate on their choice again:
\begin{itemize}
\item $p(\Mc|\info)$ uniform in the interval $1,15 \Ms$;
\item $p(\eta|\info)$ uniform in the interval $0.01,0.25$;
\item $p(\psi|\info)$ uniform in the interval $0,2\pi$;
\item $p(\iota|\info)$ proportional to $\sin\iota$ in the interval $0,2\pi$;
\item $p(t_c|\info)$ uniform in the interval $\pm$ 1 second around the correct time of arrival;
\item $p(\phi_c|\info)$ uniform in the interval $0,2\pi$;
\item $p(z,\alpha,\delta|\info)$ is set by the (3-d) positions extracted from SDSS, see Eq.~(\ref{eq:prior-z}), with the constraint $z\leq0.1$. The total number of possible hosts for each event is 362,528;
\item $p(h|\info),p(\Omega_m|\info)$ and $p(\Omega_\Lambda|\info)$ are uniform in the intervals [0.1,1.0], [0.0,1.0], [0.0,1.0], respectively.
\end{itemize}
Each of the nested sampling simulations on which the results are based has been run using a collection of 1,000 Live points providing an average of 5,000 posterior samples per GW event. For each event, the nested sampling algorithm has been run over 10 independent noise realisations. 
This totals to 10,000 simulations per detector network, yielding approximately 240,000 CPU hours per network. 

\subsubsection{Sampling the galaxy catalogue space}
For a detailed description of the Nested Sampling algorithm, the reader is refereed to \cite{Skilling:2004} for its general details and to \cite{VeitchVecchio:2010} for an implementation in the context of GW parameter estimation.
The Monte Carlo sampling of sky position and redshift is done by choosing only values of $\alpha,\delta$ and $z$ corresponding to one of the galaxies in the catalogue. However, a direct uniform sampling would be extremely computationally intensive, therefore a different scheme had to be implemented. 
At the beginning of the simulation, the full set of galaxies in the catalogue are organised in a kd-tree and each live point is assigned a triplet $\alpha_j,\delta_j,z_j$ corresponding to a random galaxy. The algorithm computes then the covariance matrix $\mathbf{C}$ for the ensemble of live points. Note that the computation of the covariance matrix is repeated every fixed number of steps of the Nested Sampling algorithm. At each iteration, when a new live point needs to replace the one having the lowest likelihood in the pool, a randomly chosen one is copied over and evolved via a MCMC procedure. Let $\alpha_0,\delta_0$ and $z_0$ be the initial values of sky position and redshift; the new values of $\alpha,\delta$ and $z$ are picked by searching the kd-tree for all galaxies within a radius $r=\sqrt{\sigma^2_\alpha+\sigma^2_\delta+\sigma^2_z}$ of the starting point $\alpha_0,\delta_0,z_0$. Among all the galaxies obtained, the proposed new live point is assigned the $\alpha,\delta$ and $z$  corresponding to a randomly picked one from the pool. Because of the very nature of the Nested Sampling algorithm, the live points tend to occupy a progressively smaller volume of the parameter space, therefore, since $\mathbf{C}$ is recomputed every fixed number of steps, the search radius $r$ decreases as the simulation progresses. With the procedure described above, the sampling of $\alpha,\delta$ and $z$ is very inefficient at the beginning of the simulation, when the diagonal elements of the covariance matrix $\mathbf{C}$ have a size comparable to the width of the prior distribution, but increases progressively. At the same, one is guaranteed an approximately uniform sampling of all galaxies.

\subsection{Single GW event}
%
%
\begin{figure*}[!ht]
\begin{tabular}{ccc}
\includegraphics[width = 2.2 in]{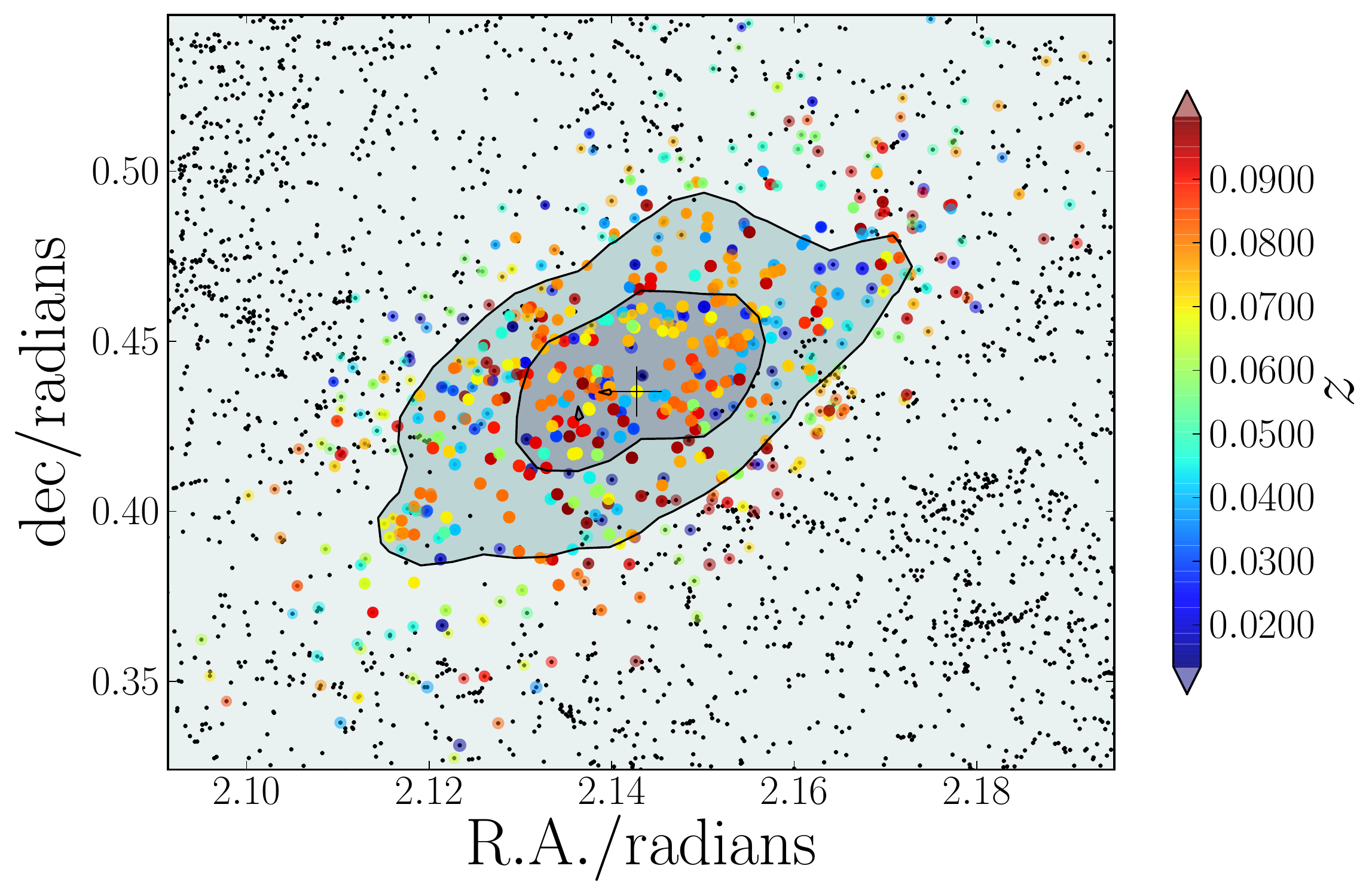} & \includegraphics[width = 2.2 in]{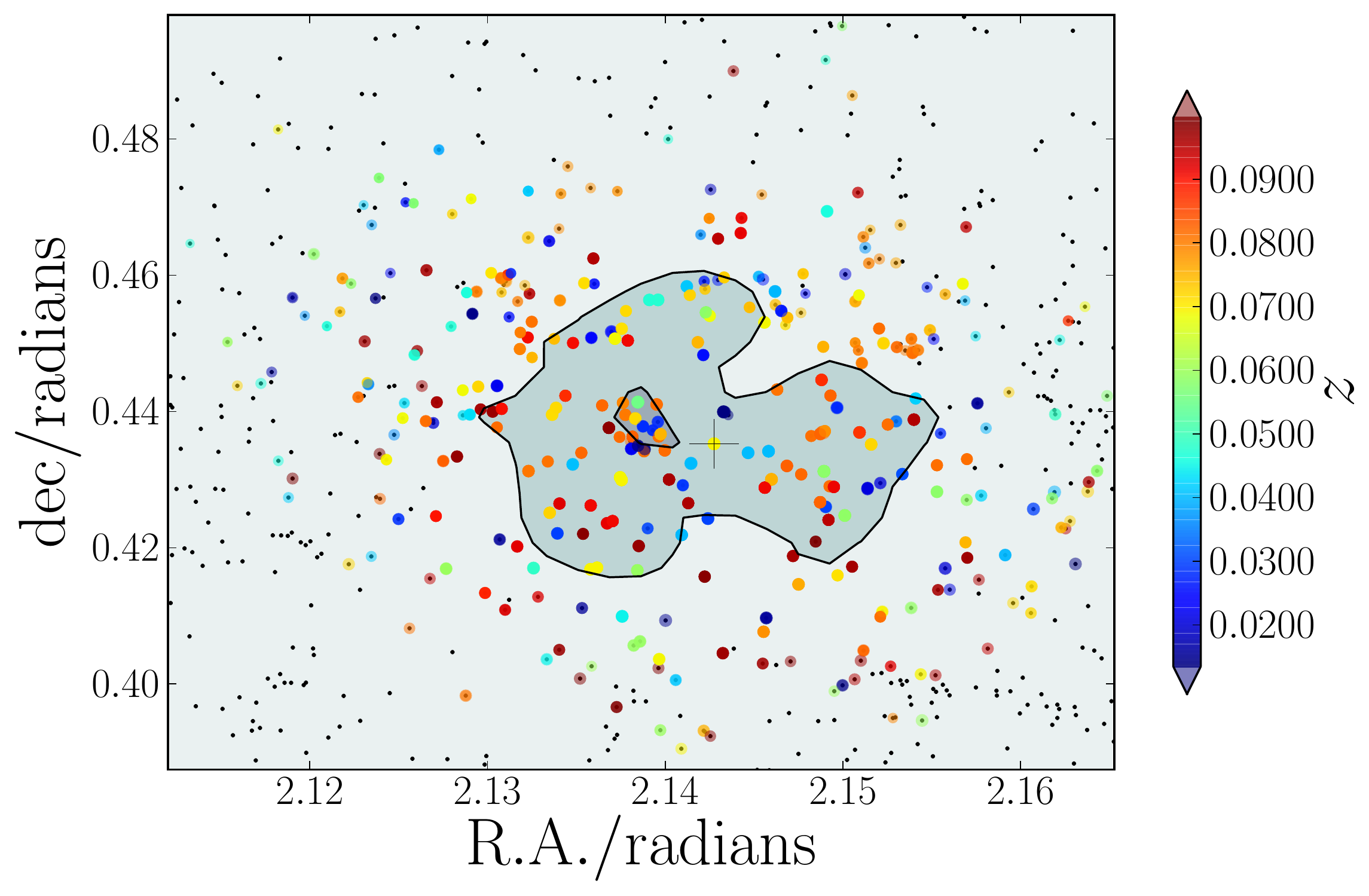} & \includegraphics[width = 2.2 in]{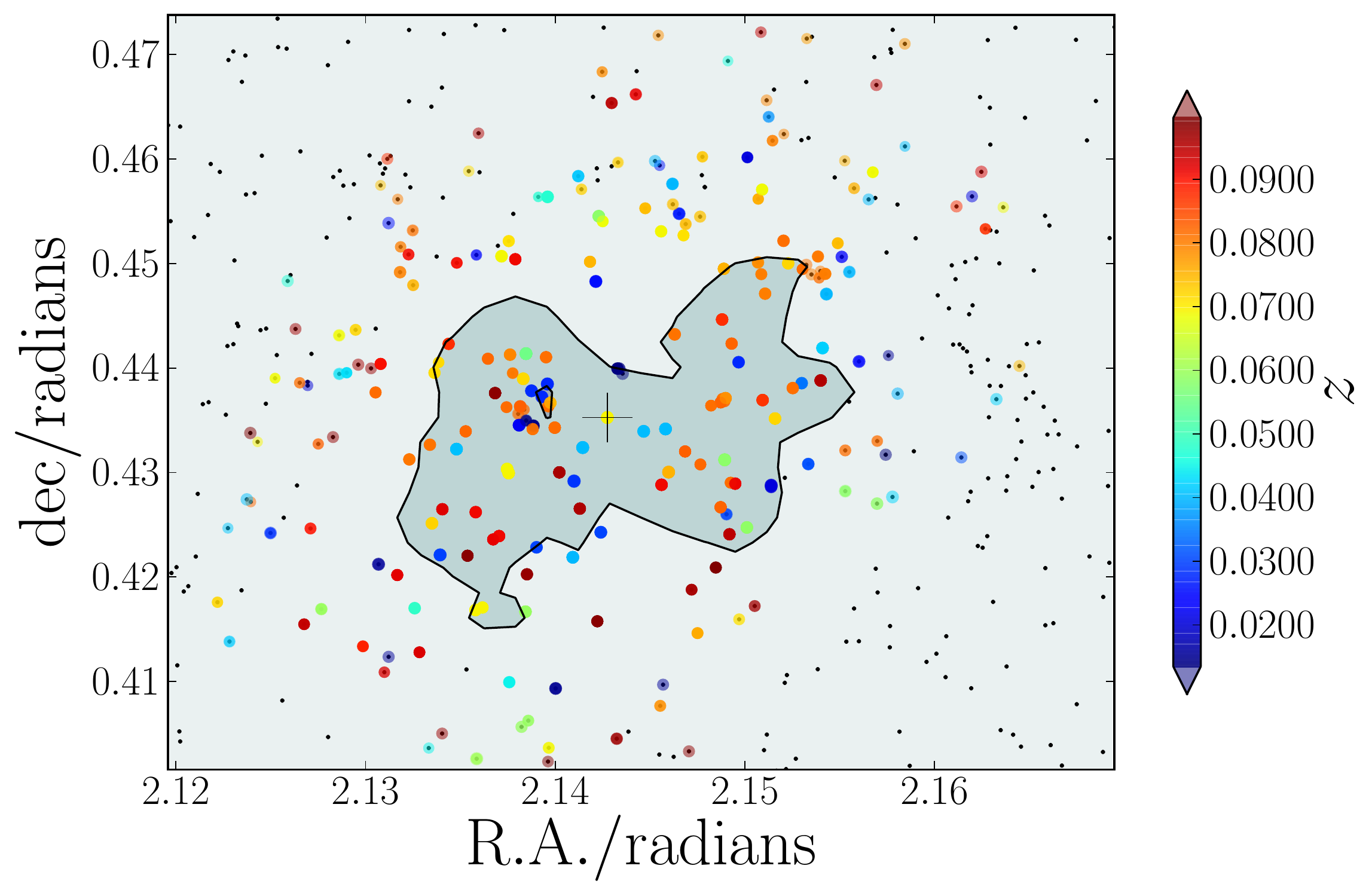}\\
\end{tabular}
\caption{Two dimensional posterior distributions for the sky position of a sample source as observed by the HLV network (left), the HLVJ network (centre) or by the HLVJI network (right). The source signal-to-noise ratios for this particular injection are H:7.1 L:7.5 V:8.5 J:7.1 I:7.0. The remaining parameters are given in Table~\ref{t:single-source}. In both panels, the cross indicates the location of the GW real host. The (coloured) dots indicate the galactic population identified as consistent with the GW event, colourcoded according to their redshift while the black dots indicate all the galaxies within the field of view.\emph{Left panel:} two dimensional posterior distribution for $\alpha$ and $\delta$ for the HLV network for which $\rho_{\mathrm{network}}\simeq13.4$. The contours indicate the 95\% and 75\% confidence intervals. The 95\% confidence area is equal to 14.8 deg$^2$ giving a total number of possible hosts of 600.  \emph{Centre panel:} two dimensional posterior distribution for $\alpha$ and $\delta$ for the HLVJ network for which $\rho_{\mathrm{network}}\simeq15.1$. The contours indicate the 95\% and 75\% confidence intervals. The 95\% confidence area is equal to 3.9 deg$^2$, within which the number of possible hosts identified is 339. \emph{Right panel:} two dimensional posterior distribution for $\alpha$ and $\delta$ for the HLVJI network for which $\rho_{\mathrm{network}}\simeq17.7$. The contours indicate the 95\% and 75\% confidence intervals. The 95\% confidence area is equal to 2.2 deg$^2$, within which the number of possible hosts identified is 230.}\label{f:sky-pos}
\end{figure*}
%
%
%
%
\begin{table*}[!ht]
\begin{tabular}{ccccccccccccc}            
\hline     
$D_L/\mathrm{Mpc}$ & $z$ & $\mathrm{dec}/\mathrm{rad}$ & $\mathrm{R.A.}/\mathrm{rad}$ & $\iota/\mathrm{rad}$ & $\psi/\mathrm{rad}$ & $\Mc/\Ms$ & $\eta$ & $\rho_\mathrm{H}$ & $\rho_\mathrm{L}$ & $\rho_\mathrm{V}$ & $\rho_\mathrm{J}$ & $\rho_\mathrm{I}$ \\
\hline
$313$ & $0.069381$ & $0.435262$ &  $2.142747$ & $0.339614$ &  $0.519744$ & $6.350444$ & $0.178603$ & $7.1$ & $7.5$ & $8.5$ & $7.1$ & $7.0$ \\ 
\hline                              
\end{tabular}
\caption{Summary of the properties of the source to which the results presented in the subsection refer.}
\label{t:single-source}
\end{table*}
%
%
%
%
\begin{figure*}[!ht]
\begin{tabular}{ccc}
\includegraphics[width = 2.2 in]{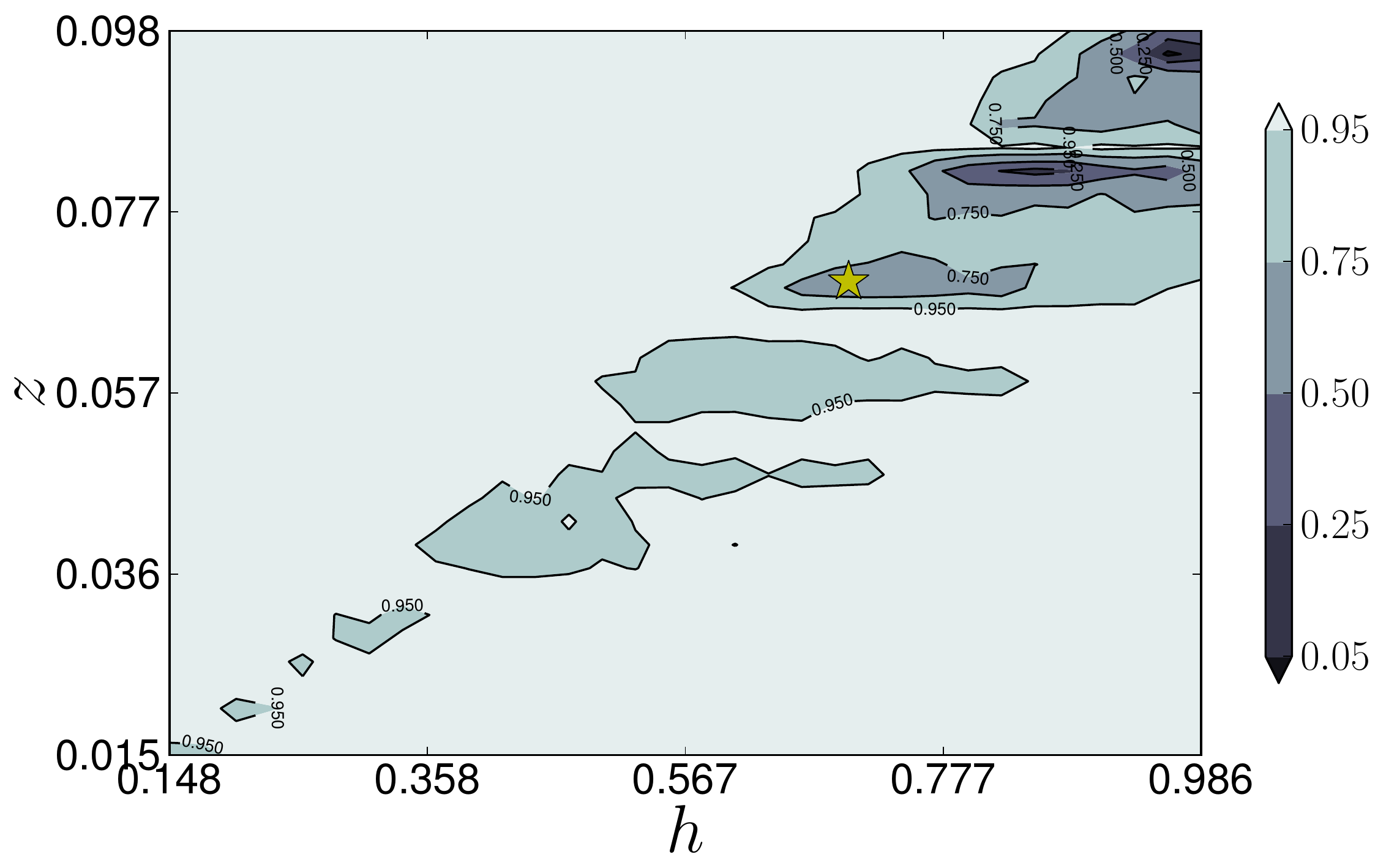} & \includegraphics[width = 2.2 in]{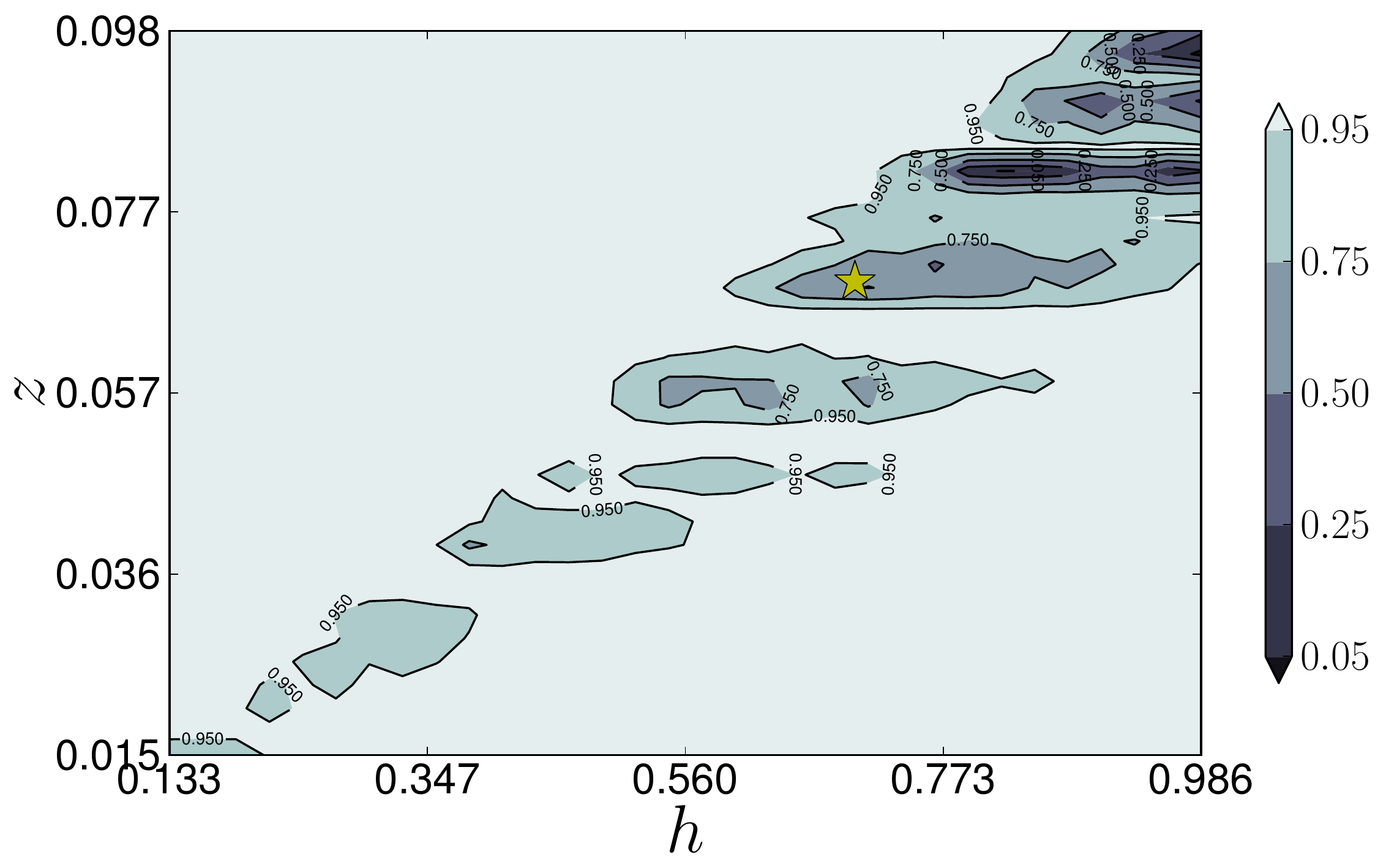} & \includegraphics[width = 2.2 in]{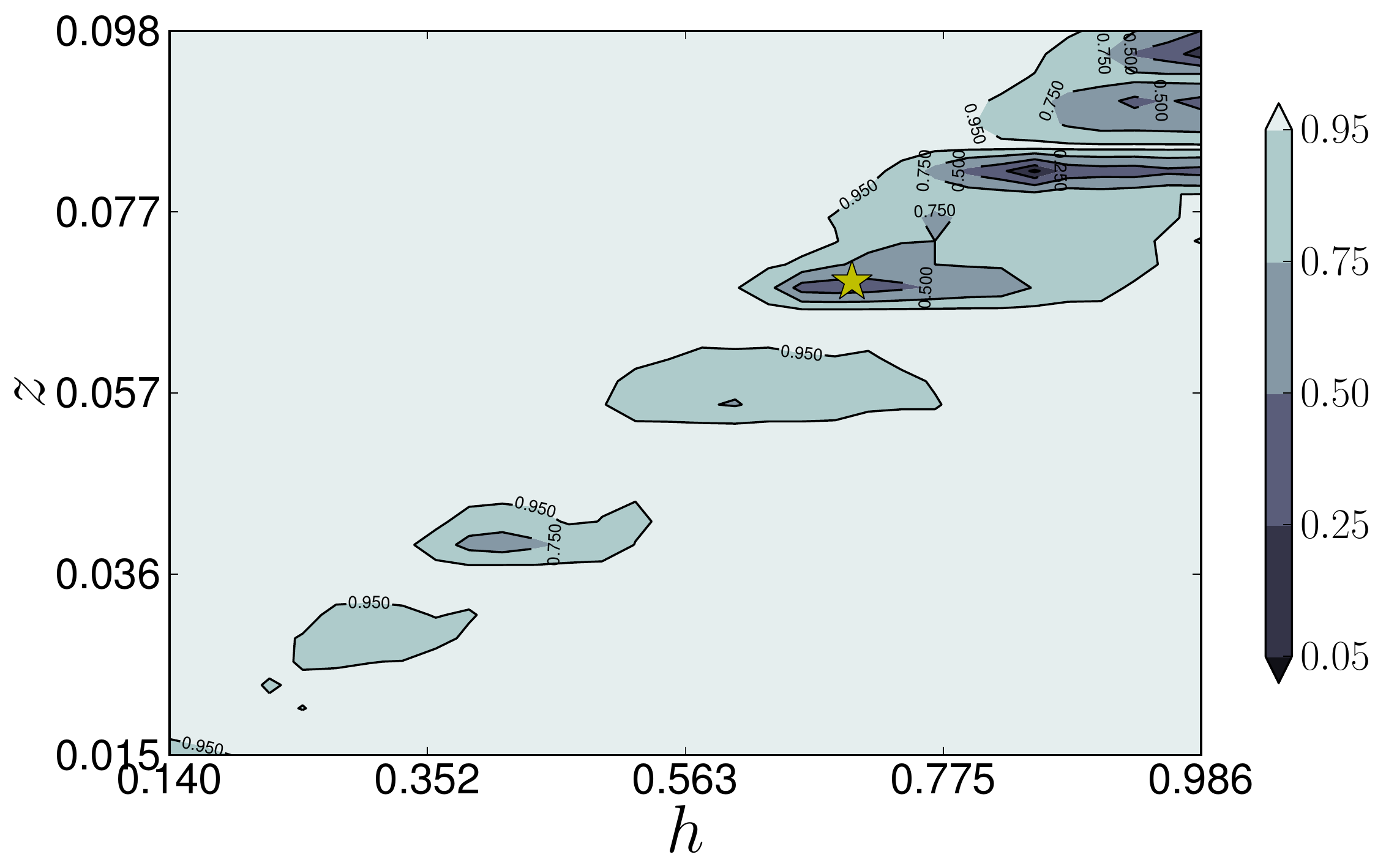}\\
\includegraphics[width = 2.2 in]{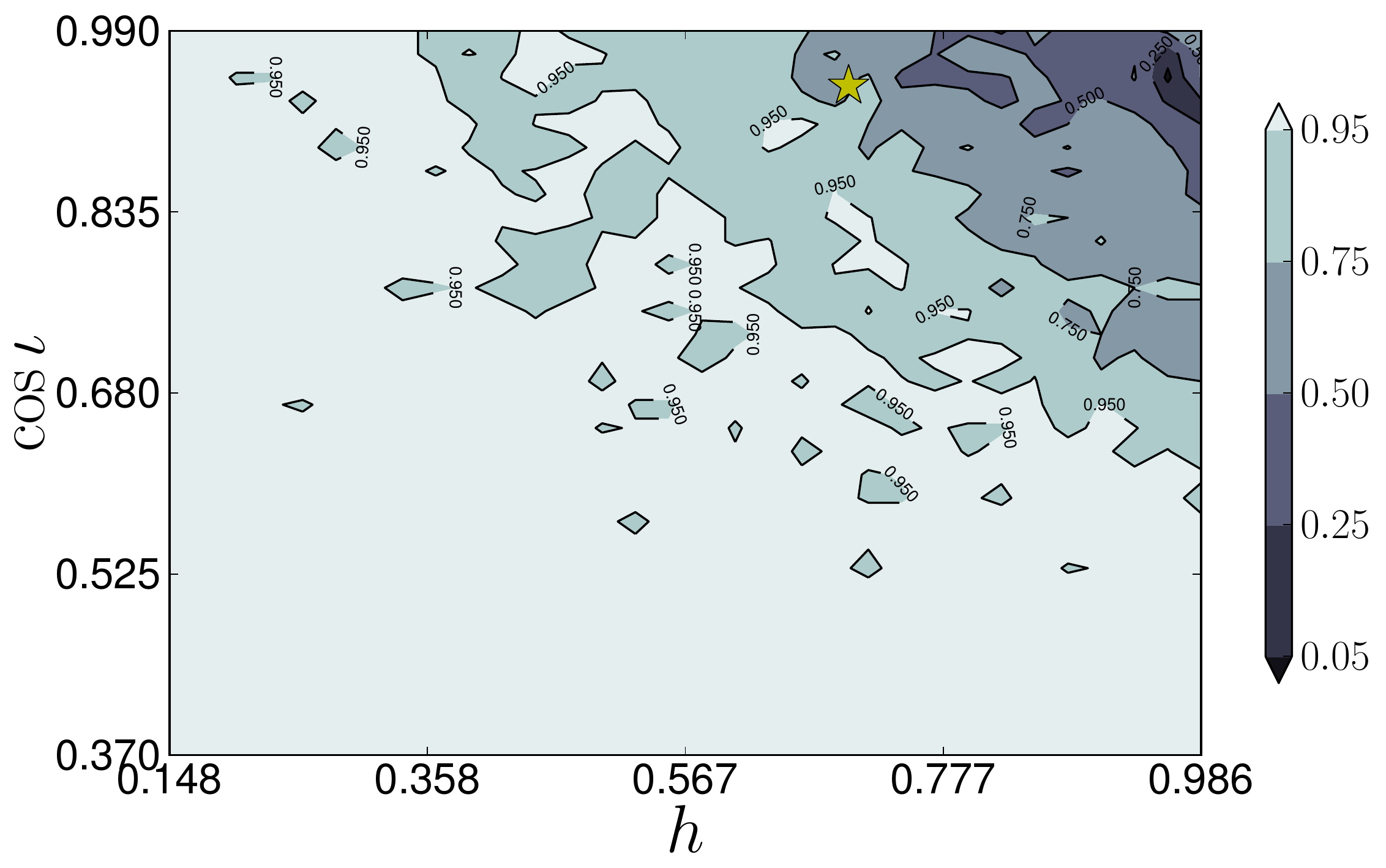} & \includegraphics[width = 2.2 in]{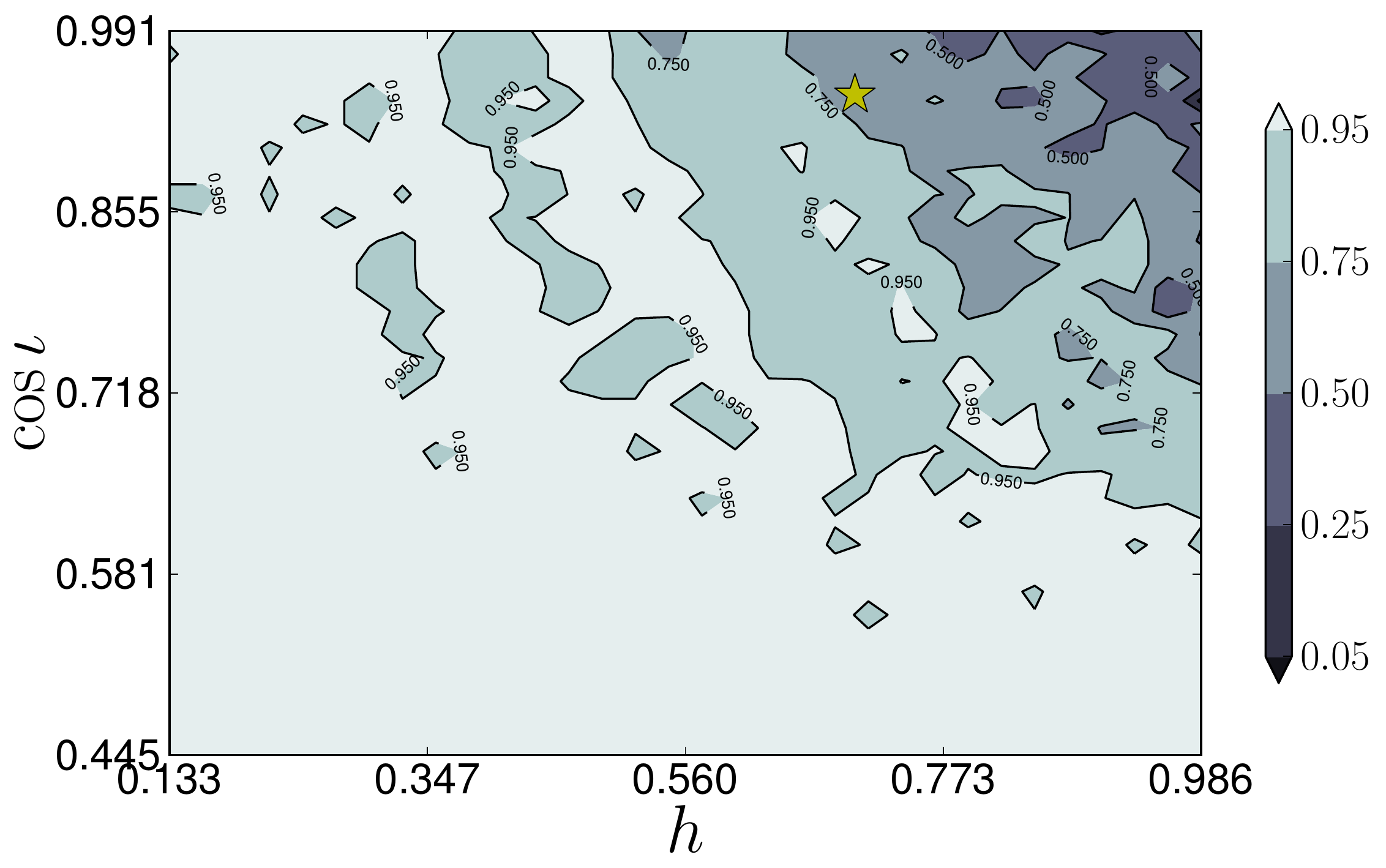} & \includegraphics[width = 2.2 in]{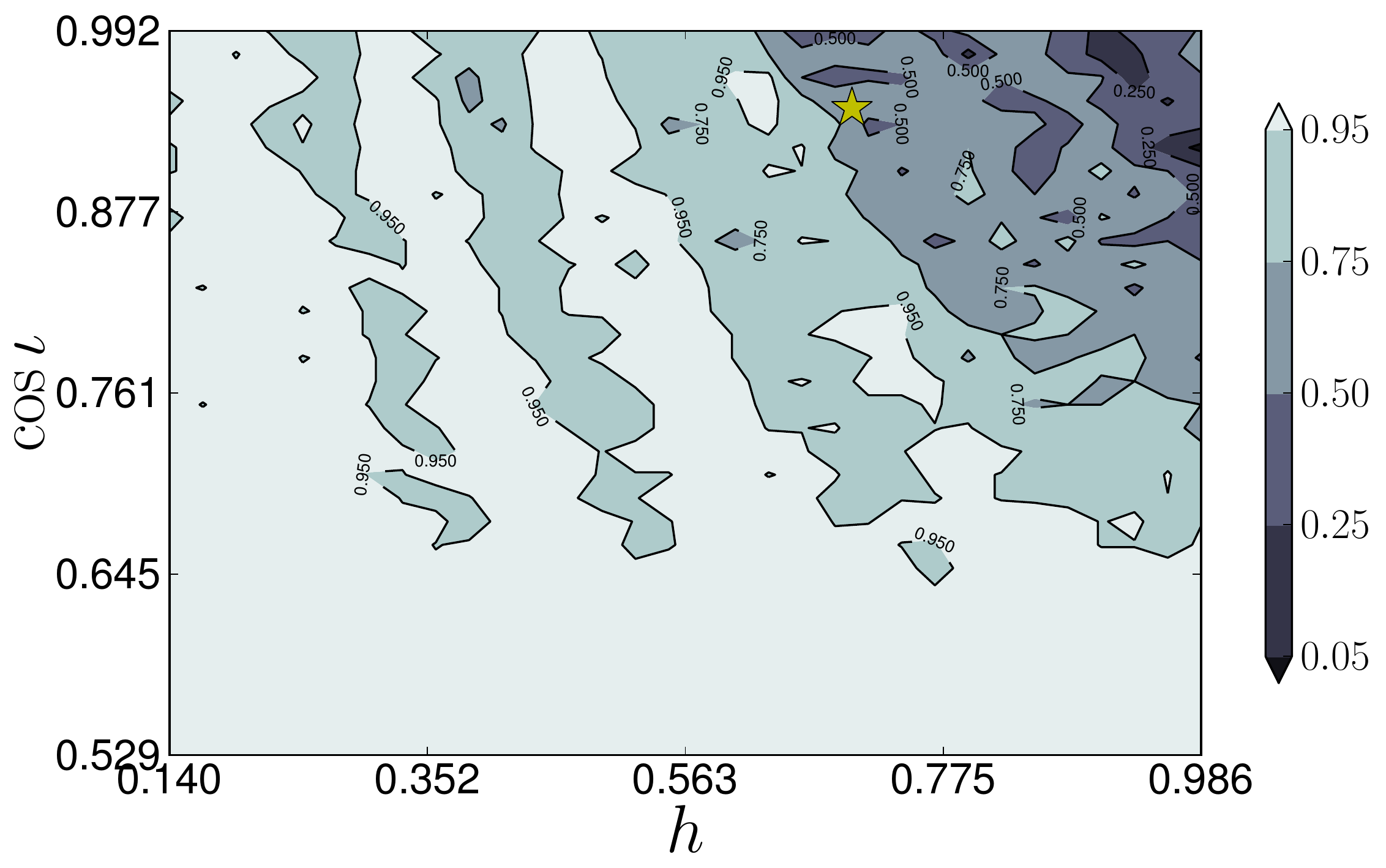}\\
\end{tabular}
\caption{\emph{Top panels}: joint two dimensional posterior distributions for the redshift and for $h$ of the same source as in Fig.~\ref{f:sky-pos} as observed by the HLV network (left), the HLVJ network (centre) or by the HLVJI network (right). The star indicates the indicates the real value of the redshift and of $h$. \emph{Bottom panels}: joint two dimensional posterior distributions for $h$ and $\cos\iota$. The star indicates the injection value. In all panels, the contours  indicate the 95\%, 75\%, 50\%, 25\% and 5\% confidence intervals.
In all cases it is evident the strong correlation between $h$ and $z$, which is an obvious consequence of Eq.(\ref{eq:dl}), and between $h$ and $\cos\iota$. This last degeneracy is just the translation of the known $D_L$--$\iota$ degeneracy that ultimately is the limiting factor in the determination of the parameters appearing in the amplitude of the GW. Moreover, the distributions are multimodal, corresponding to the different combinations of $h, z$ and $\cos\iota$ that give constant $D_L$. The accuracy of the estimation of the redshift is similar for all networks. However, it is noticeable the increase in resolving power for $\cos\iota$ when more detectors are considered.}\label{f:z-pos}
\end{figure*}
%
%
Without a one-to-one electromagnetic identification, from a single GW event it is not possible to measure $\cosmoP$ with sufficient accuracy. For a reasonable estimate, it is pivotal the combination of the information coming from a number of sources \cite{Schutz:1986}. However, it is still interesting to quantify the performance of the networks under consideration and compare them. As it is impossible to report a detailed comparison for all 1,000 GW events, what follows will concentrate on the direct comparison of a particular GW event chosen from the mode of the signal-to-noise ratio distribution shown in Fig.~\ref{f:catalog} and then report a few statistical properties for the whole 1,000 events, Table~\ref{t:single-statistics}. The parameters of the sample event under consideration are given in Table~\ref{t:single-source}. Fig.~\ref{f:sky-pos} shows the joint two dimensional posterior distribution for $\alpha$ and $\delta$ for the two networks under consideration for the source whose parameters are given in Table \ref{t:single-source}. The large dots are the galaxies identified as the possible hosts of the GW event. The first striking difference between the HLV and the remaining two networks is the sky resolution. The benefits of adding more interferometer to the global network has been already strongly stressed by the works in \cite{NissankeEtAl:2010,VitaleZanolin:2011,Schutz:2011,VeitchEtAl:2012}, here we see how a better localisation accuracy translates in the number of identified putative hosts. For the same event considered as an example in this section, the 95\% confidence area goes from 14.8 deg$^{2}$ measured by the HLV network to 3.9 deg$^2$ for the HLVJ one and to 2.2 deg$^2$ instead for the HLVJI one. This translates in a number of hosts identified of 600, 339 and 230, respectively.
However, the redshift $z$ and $h$ are inferred with similar accuracies, see Fig.~\ref{f:z-pos} for $h$ and $z$. 
The joint $z$ and $h$ posterior shows some interesting features; from the two dimensional distribution we evince the multimodal character of the distribution itself. The comparison with the joint posterior for $\cos\iota$ and $h$ sheds some light on the nature of the multimodality; the amplitude of a GW in fact depends mostly on the instrument geometrical response to the wave and its distance, since the chirp mass is estimated from phase information. Since $\alpha$ and $\delta$ are determined mainly by the relative phase shifts in each instruments, in other words by the different times of arrival of the GW at each detector location, the only angular variables that affect the amplitude determination are the orientation of the binary $\psi$ and $\iota$. The polarisation $\psi$ is determined by the antenna pattern function once $\alpha$ and $\delta$ have been constrained and what is left affecting the amplitude of the GW is only $\iota$. In fact, Fig.\ref{f:z-pos} bottom panels shows clearly the correlation between $h$ and $\cos\iota$. There is, however a remarkable difference between the three networks. In the three interferometers case the $\cos\iota,h$ distribution is not multimodal. The reason for this has been already given implicitly above; the HLV network is very much less accurate in the determination of the sky position of a source, therefore the constraints from the geometrical response to the GW are much weaker. This implies that neither $\psi$ or $\cos\iota$ are as well constrained which leaves more freedom to $z,\cos\iota$ and $h$ to rearrange and give the same value of the likelihood. This further stresses the importance of combining multiple observations to break this 3-way degeneracy if we are to estimate $\cosmoP$ at all from GW. 

\begin{table*}[!ht]
\begin{tabular}{c|cccccccccccccc}            
\hline     
network & $h$ & $\Omega_m$ &$\Omega_k$ & $\Omega_\Lambda$ & $z$ & $\mathrm{dec}/\mathrm{rad}$ & $\mathrm{R.A.}/\mathrm{rad}$ & $\cos\iota$ & $\psi/\mathrm{rad}$ & $t_c/\mathrm{ms}$ & $\Mc/\Ms$ & $\eta$ & $N_{\mathrm{galaxies}}$ & $\sqrt{N_{\mathrm{galaxies}}^2- \langle N_{\mathrm{galaxies}}\rangle^2}$\\
\hline
HLV & $0.63$ & $0.95$ & $1.55$ &  $0.95$ & $0.04$ &  $0.05$ & $0.05$ & $0.45$ & $1.6$ & $1.1$ & $0.01$ & $0.01$ & $283$ & $332$\\
HLVJ & $0.57$ & $0.95$ & $1.55$ &  $0.95$ & $0.04$ &  $0.04$ & $0.03$ & $0.40$ & $1.6$ & $0.6$ & $0.007$ & $0.01$ & $171$ & $192$\\
HLVJI & $0.54$ & $0.95$ & $1.55$ &  $0.95$ & $0.03$ &  $0.03$ & $0.02$ & $0.34$ & $1.6$ & $0.4$ & $0.006$ & $0.01$ & $118$ & $137$\\
\hline                              
\end{tabular}
\caption{Noise averaged median 95\% confidence intervals widths from the 1,000 GW events for all the parameters measured and for the three networks under consideration. The last two columns report the median number of galaxies identified as potential hosts and its variance, respectively. Please note that the 95\% confidence width for $\psi$ is not a reliable indication of the performance of the networks since the posterior distribution for $\psi$ is typically bimodal and the modes are spaced by $\pi$.}
\label{t:single-statistics}
\end{table*}
Table \ref{t:single-statistics} reports the median 95\% confidence intervals widths for all the parameters estimated by each simulation for all the GW events observed from the HLV, top row, the HLVJ, central row, and HLVJI, bottom row, networks. None of the networks is able to provide, on average, a clean measurement of any of the $\cosmoP$. The median 95\% interval widths for $\Omega_m$ and $\Omega_\Lambda$ are equal to 0.95, therefore we expect that these two parameters will not be measurable as this is also the 95\% width of their prior distributions. So, none of the networks will be able to constrain the energy density parameters $\Omega_m$ and $\Omega_\Lambda$. This is not surprising as, for $z\simlt 0.1$, a change in one of the energy density parameters is reflected in a change in $D_L$ by about 1\%, which is very much smaller than the typical measurement uncertainty on $D_L$ of $\simgt 30\%$. However, the median 95\% width for $h$ is 0.63--0.54, therefore, on average, the 95\% width of the posterior distribution is about half the size of the prior width. $H_0$ can be measured by second generation interferometers. In the next section, we will find out to which accuracy this measurement can be done. 
Regarding the other parameters, the findings presented in Table \ref{t:single-statistics} confirm and are in agreement what was already found by several other studies \cite{NissankeEtAl:2010,VitaleZanolin:2011,Schutz:2011,VeitchEtAl:2012}. A fourth detector in the world wide network will improve the sky localisation accuracy by a factor of $\sim 2$ and the advantage in having more than four detectors is marginal. This improvement is mostly due to the more accurate determination of the time of coalescence $t_c$ from the higher number of relative time delays between the detectors. A more precise determination of the position of a source on the celestial sphere has very important consequences. The obvious ones for the electro-magnetic follow-up of GW observations are discussed in \cite{Schutz:2011}. However, there are more subtle consequences of a better sky localisation. When a galaxy catalogue is used as a prior, what one obtains is a set of galaxies that due to their position in the sky are classified as potential hosts of the current GW event. For each event then the properties of the putative galactic population can be studied statistically. For example one can study the luminosity function of the potential hosts or their clustering. This kind of studies would indicate which morphological types are more likely to host compact coalescing binary and their typical masses and colours as well as the properties of their environment. Having a smaller number of putative counterparts ensures a faster emergence of the features that characterise the typical GW event host. What can be learnt from the galaxy population will be the object of future studies.  
As found in \cite{VeitchEtAl:2012}, the measurement of the inclination is improved by a factor of $\sim 10$\% in going from three to four detectors and marginally in going from four to five. The redshift of a source, when using a galaxy catalogue as prior as in this study, can be determined with essentially the same accuracy, regardless of the number of detectors constituting the GW network. 

\subsection{Multiple GW events}

\begin{figure*}[!ht]
\begin{tabular}{ccc}
\includegraphics[width = 2.2in]{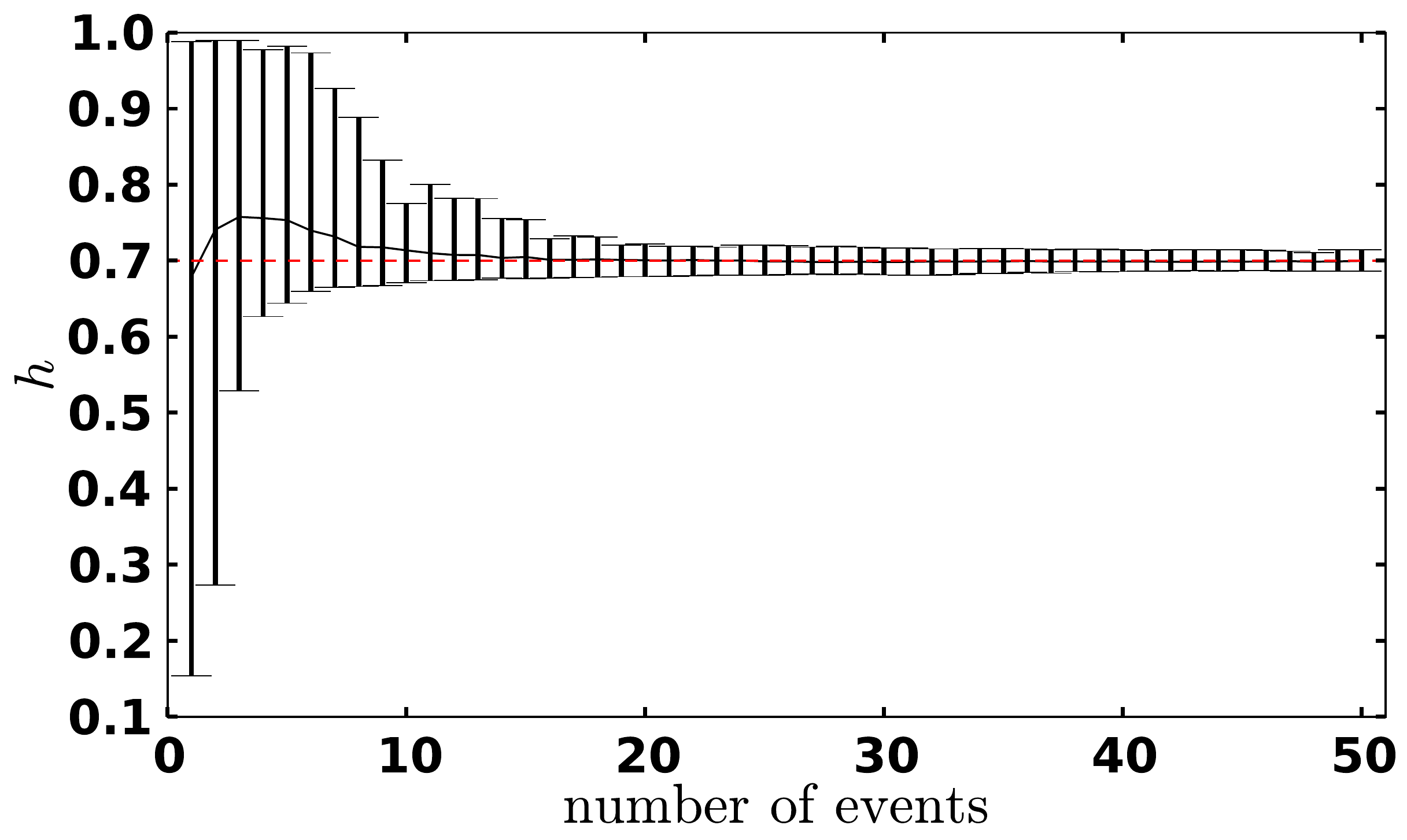} & \includegraphics[width = 2.2in]{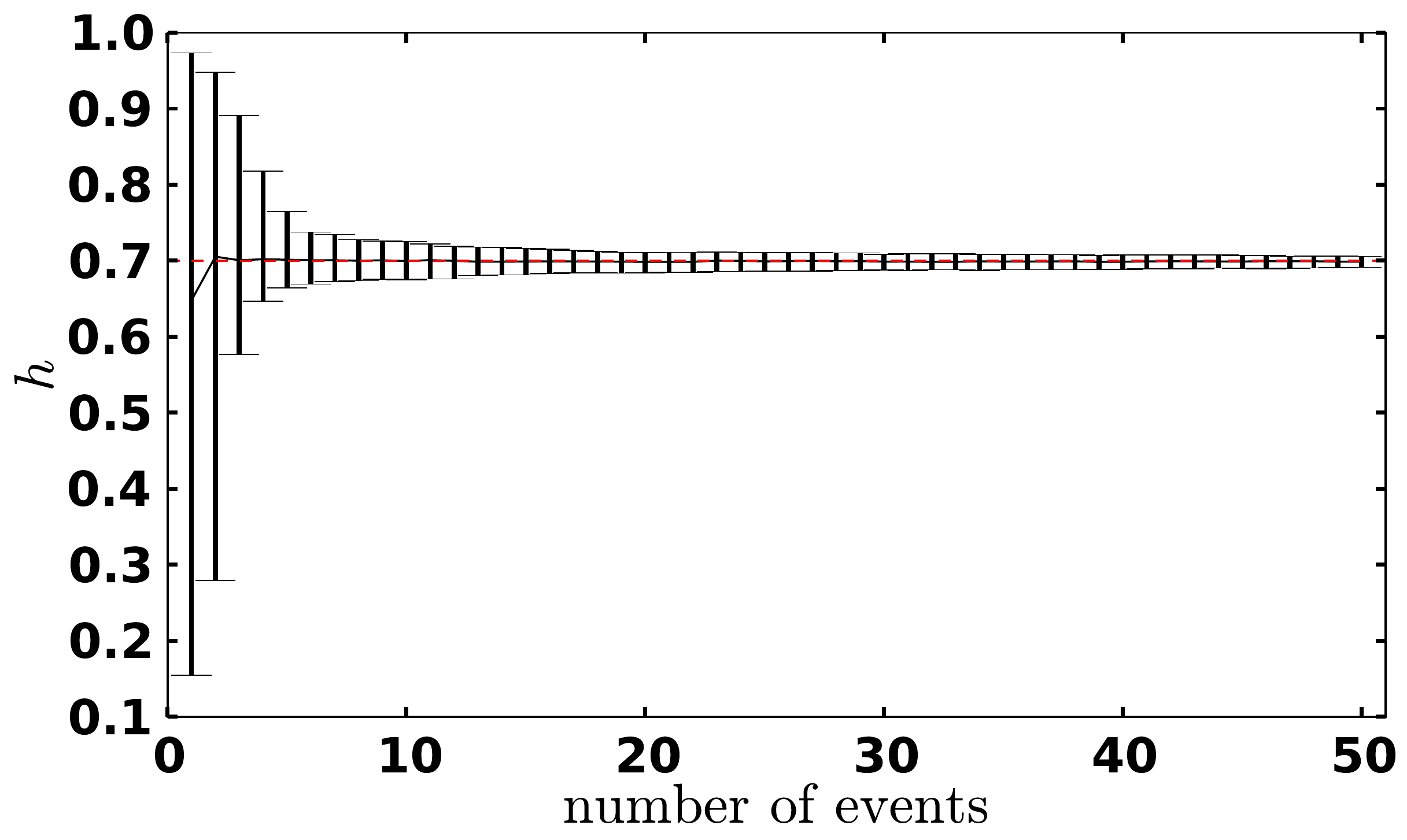} & \includegraphics[width = 2.2in]{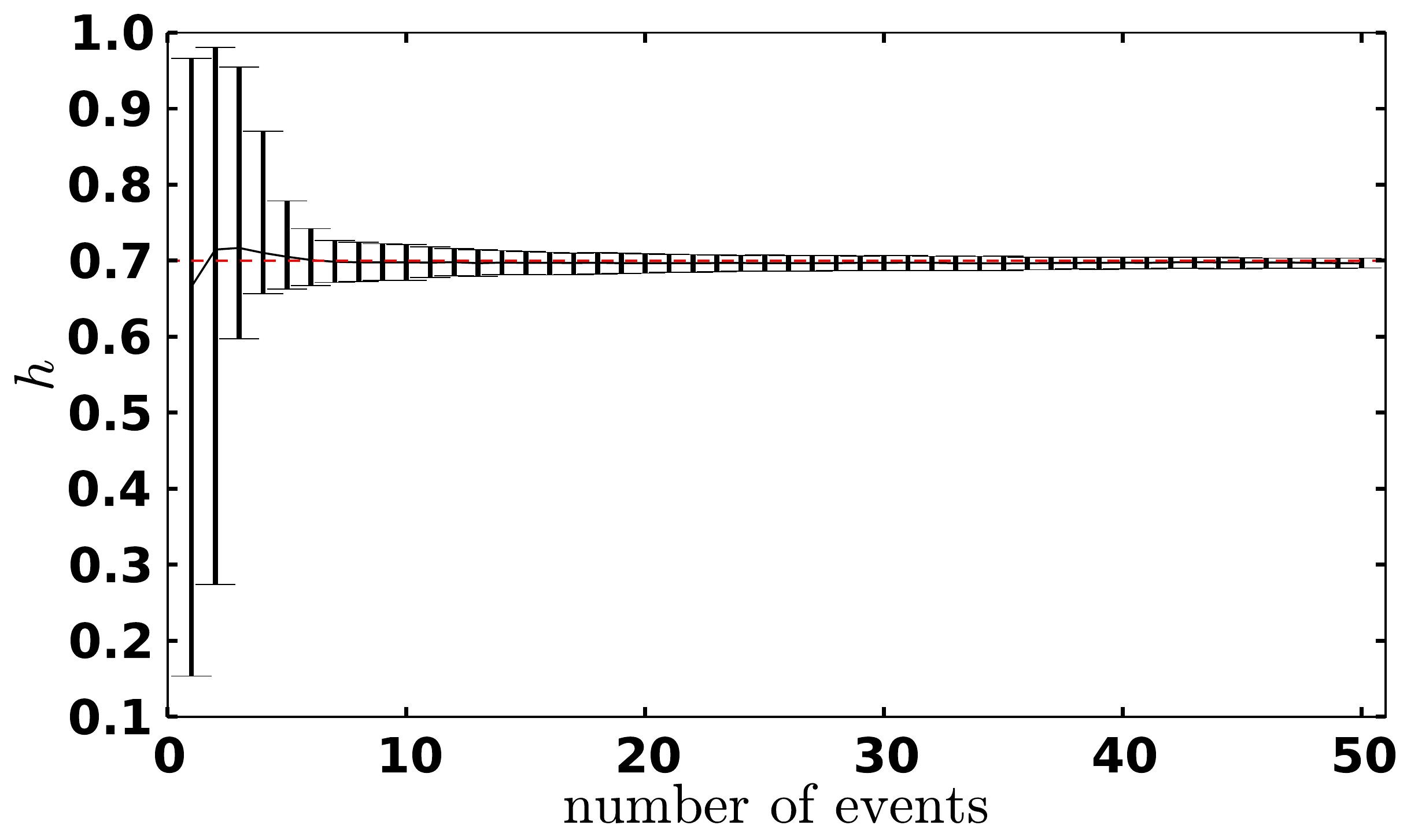}\\
\includegraphics[width = 2.2in]{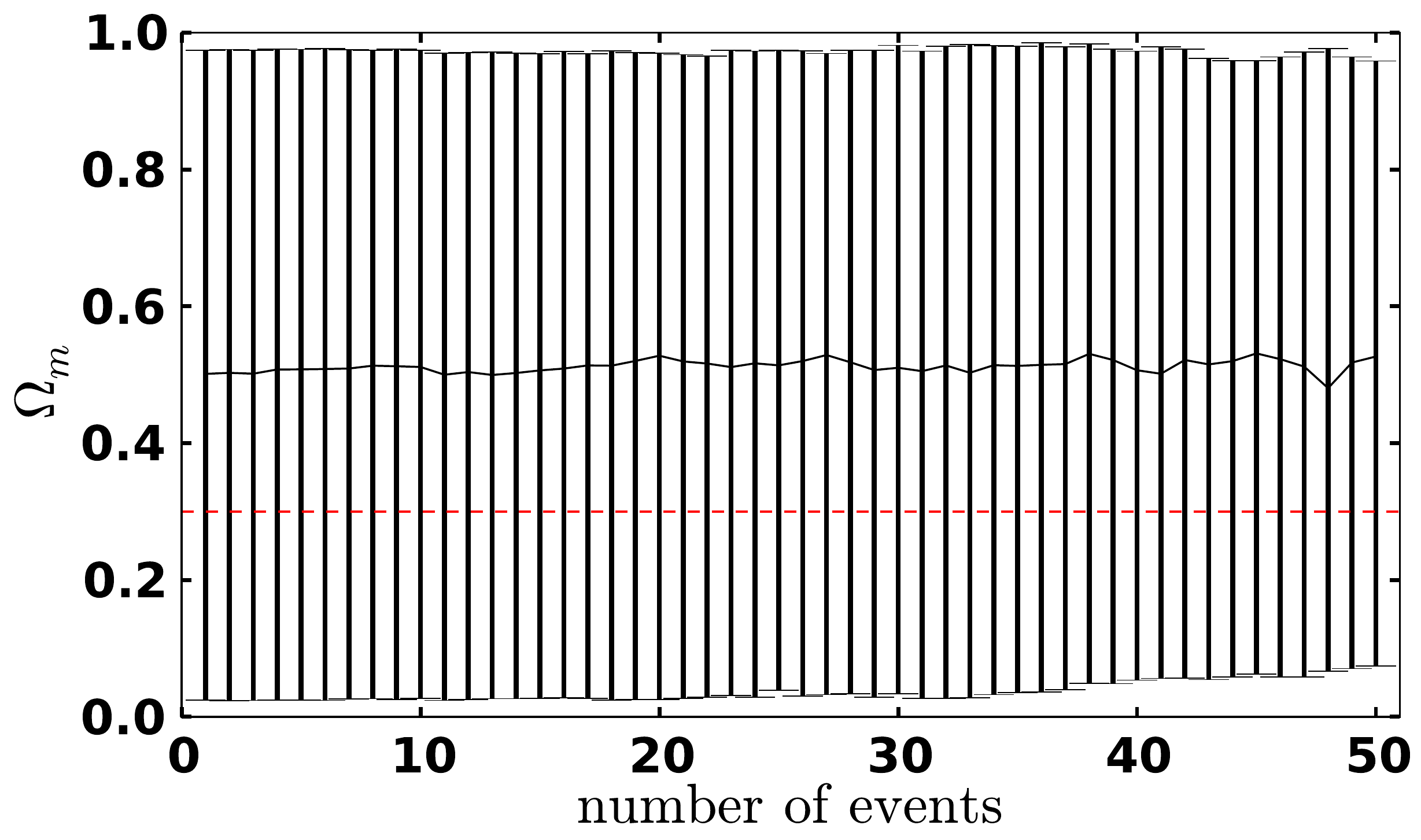} & \includegraphics[width = 2.2in]{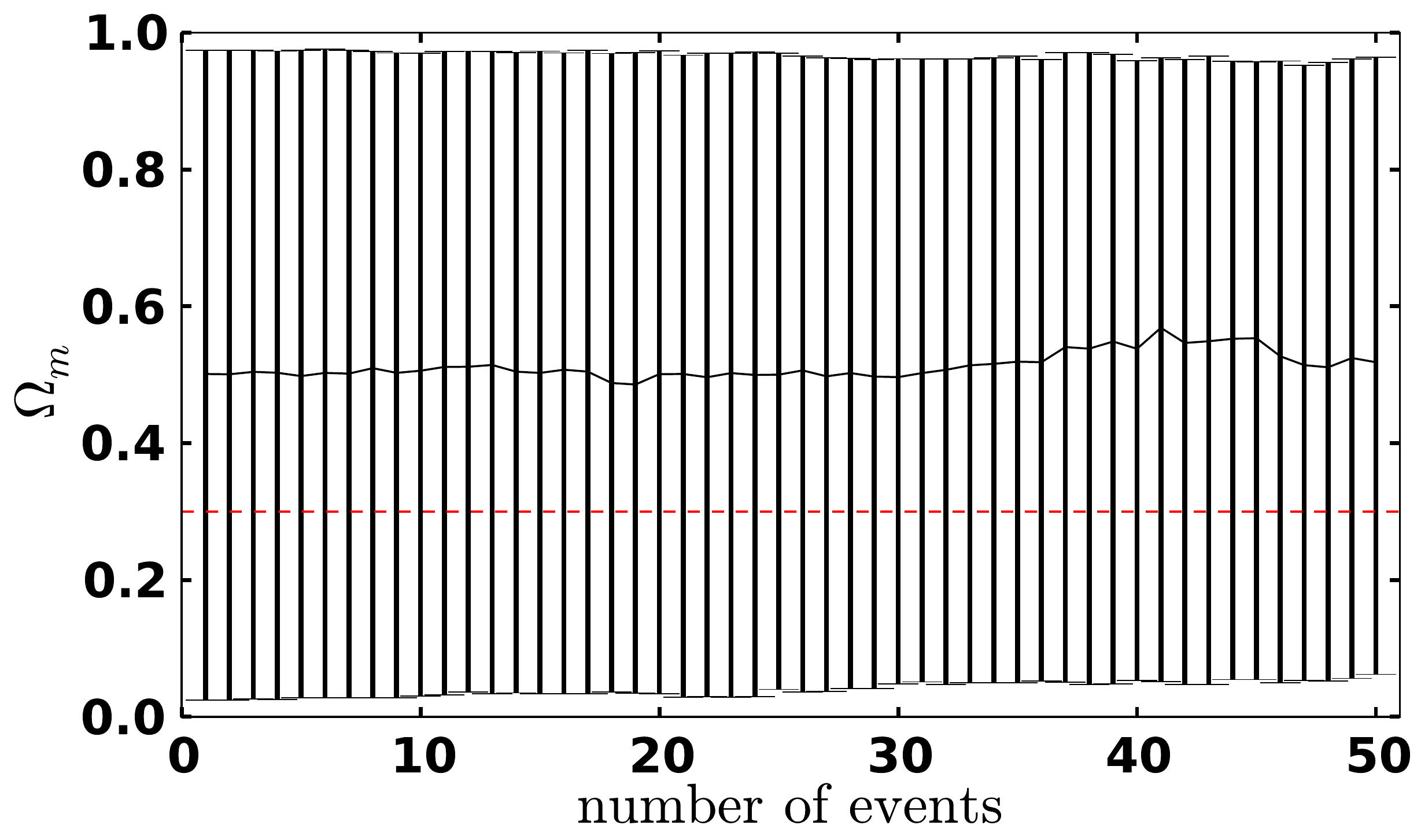} & \includegraphics[width = 2.2in]{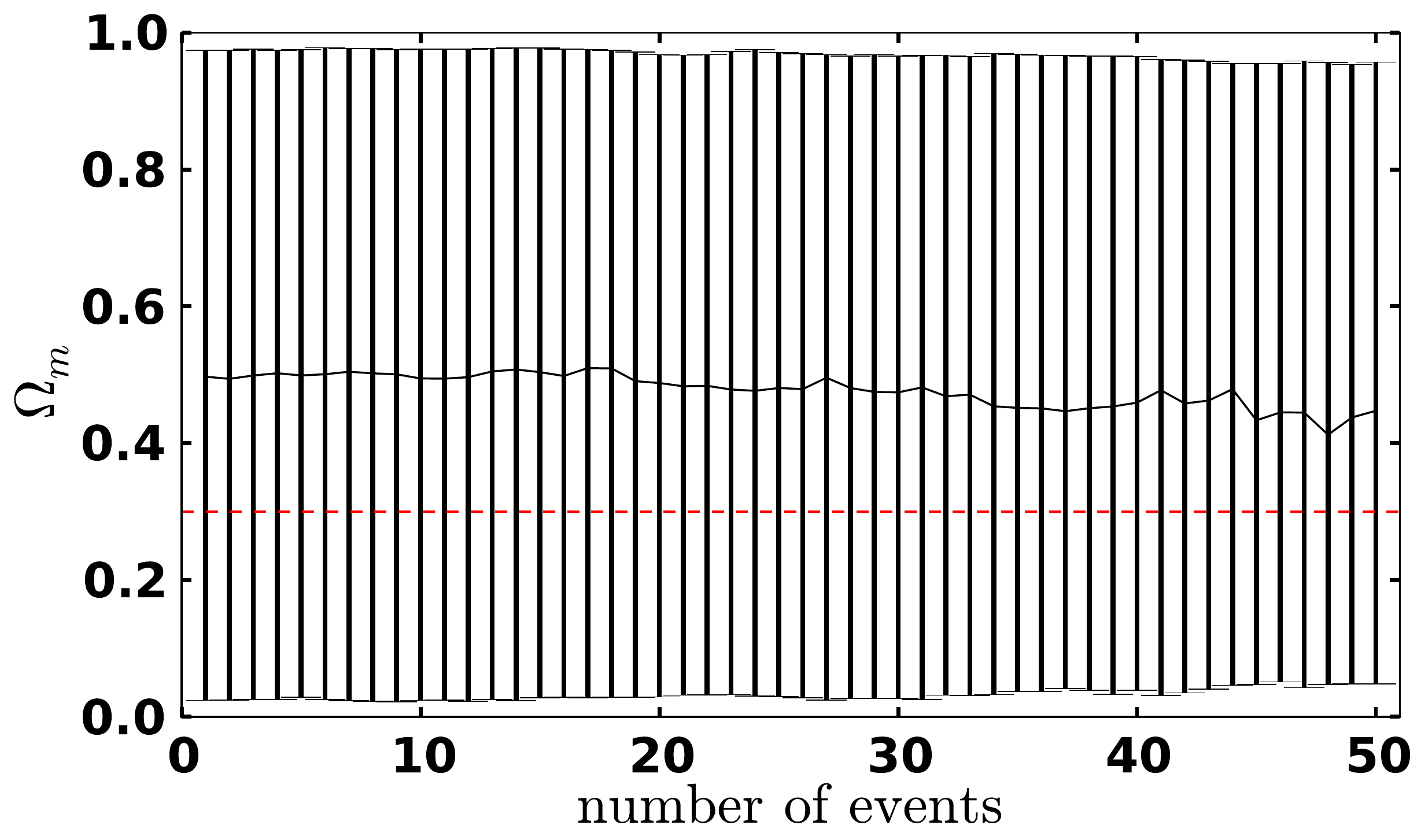}\\
\includegraphics[width = 2.2in]{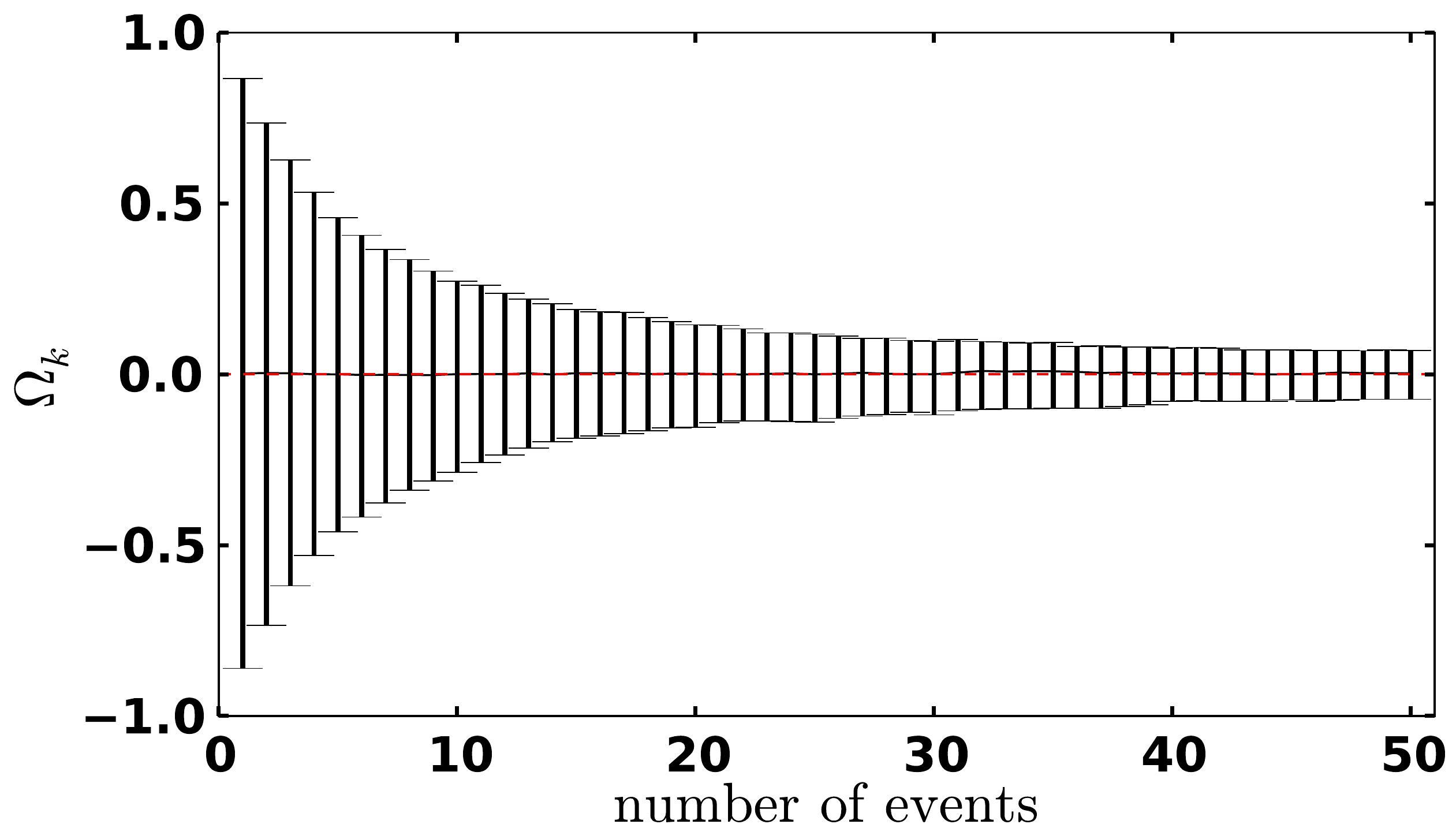} & \includegraphics[width = 2.2in]{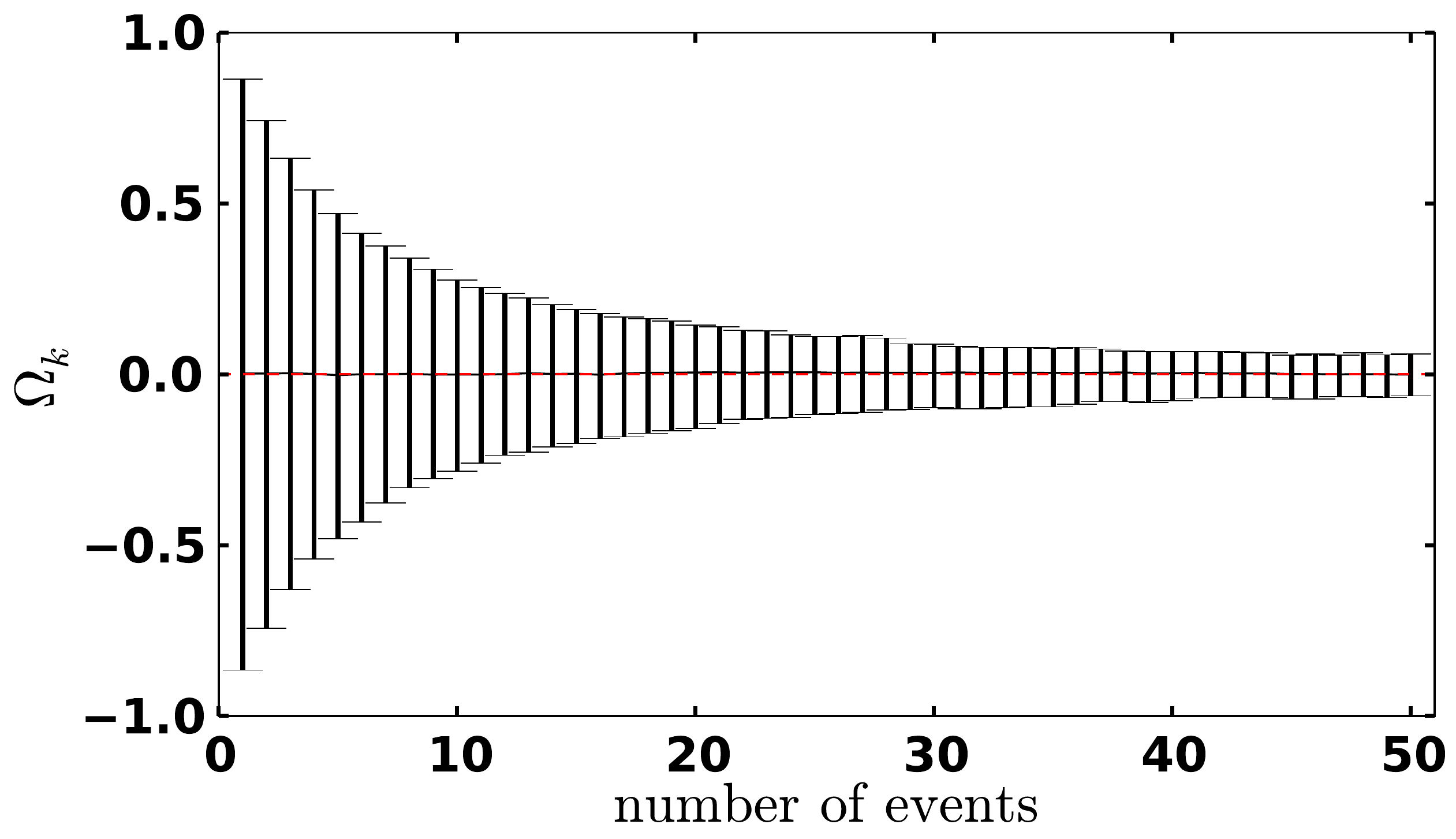} & \includegraphics[width = 2.2in]{Ok_HLVJI.pdf}\\
\includegraphics[width = 2.2in]{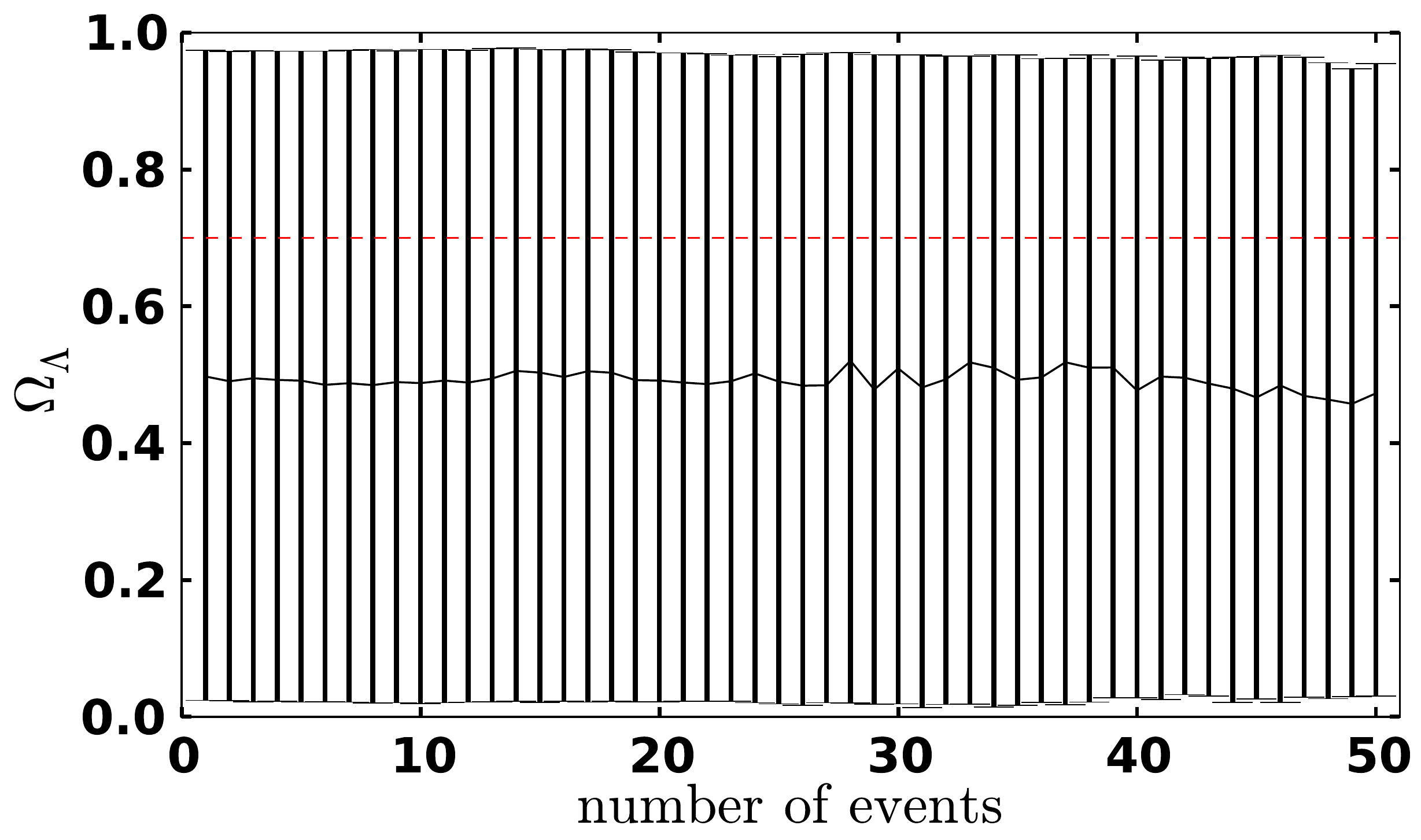} & \includegraphics[width = 2.2in]{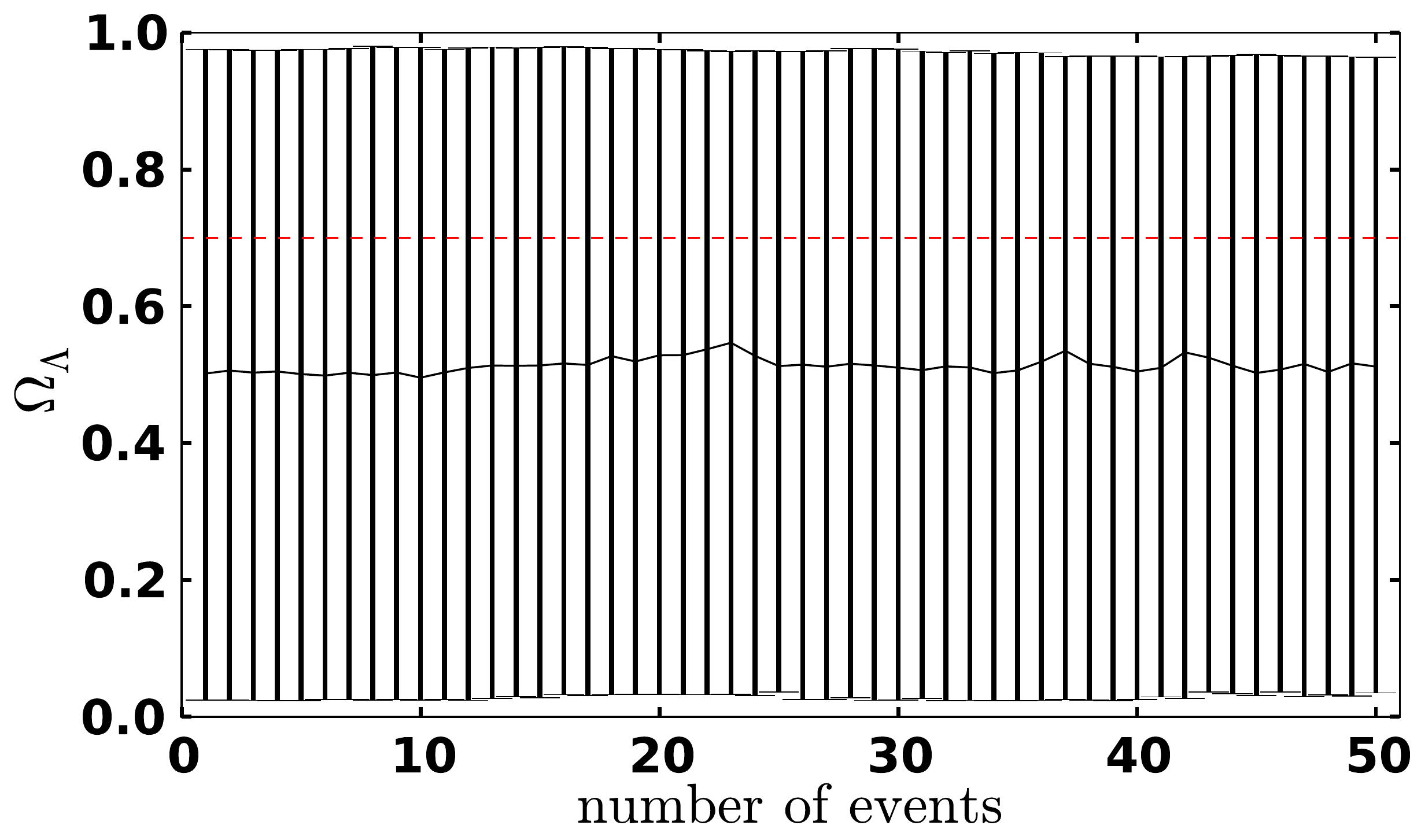} &\includegraphics[width = 2.2in]{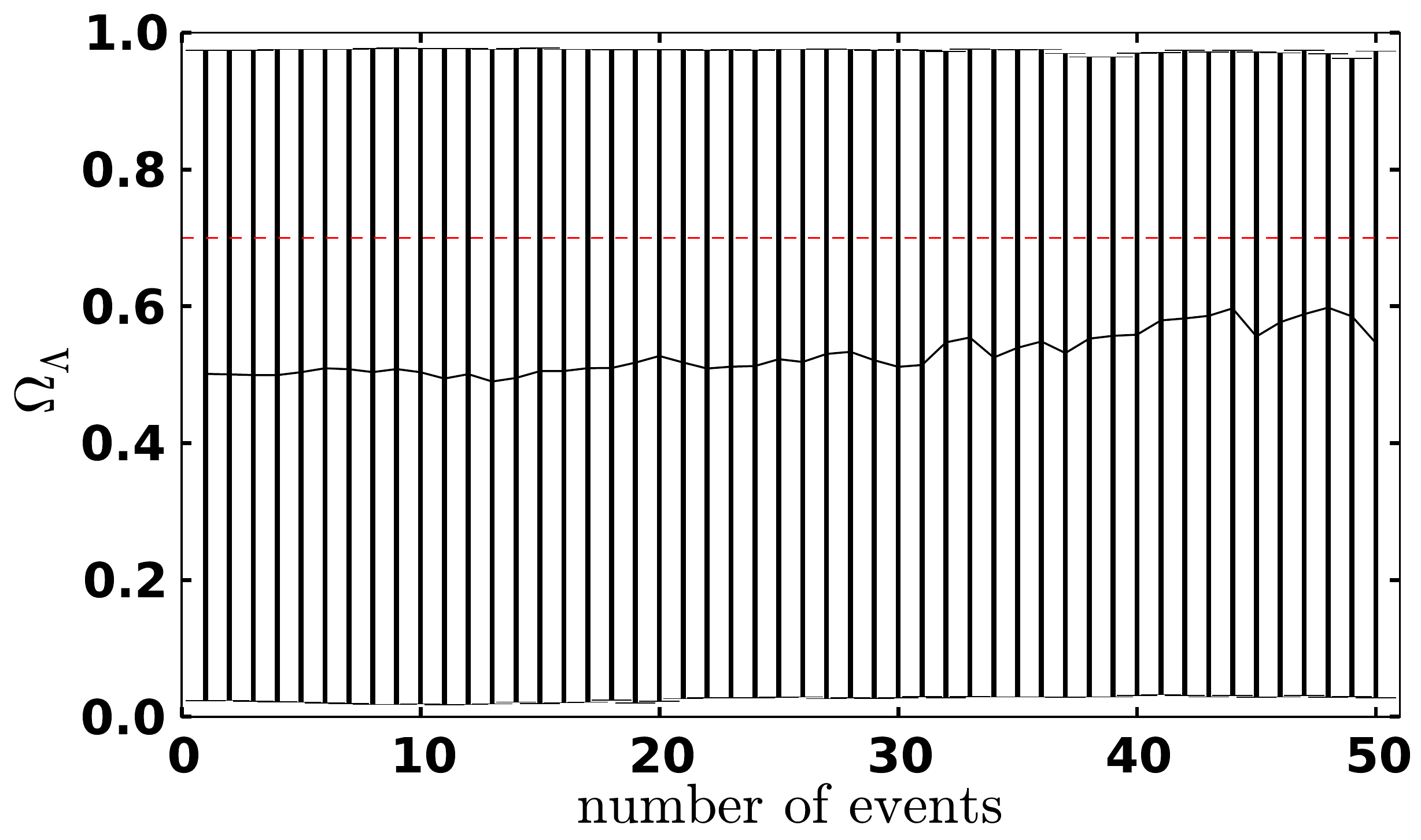}\\
\end{tabular}
\caption{Evolution of the medians (solid line) and 95\% confidence intervals evolution for $\cosmoP$ as a function of the number of events included in the computation of the joint posterior distributions. Each data point, and relative error bar, is the result of averaging over 20 independent realisations of the GW events catalogue. The left columns presents the result for the HLV network, the center column for the HLVJ network while the right column is relative to the HLVJI network. In all figures the (red) dashed line is the injected value. The irregular jumps are due to the finite bin size. The convergence of $\Omega_k$ towards the injected value 0 is artificial and due to the prior probability choice for the analysis. See text for a discussion.}\label{f:combined-pos}
\end{figure*}

We have already stressed many times the importance of combining the information coming from multiple GW events for the purpose of inferring the value of $\cosmoP$. In this subsection I will present the result of computing the joint posterior distribution on $\cosmoP$ using Eq.~(\ref{eq:posteriors-multi-events}). In particular, the results will be presented in the form of an average over 20 independent GW events catalogues obtained from the 1,000 events presented in section \ref{s:catalog}. The joint posterior distributions have been computed by histogramming the posterior samples from each Nested Sampling chain and then combined together using the correspondence between an histogram and the Dirichlet distribution that describes the probability of each sample to end up in a particular bin \cite{DelPozzoEtAl:2011}. This approach avoids procedures like convolving with a Gaussian kernel, which might smooth out features of the distribution, and having zeros in any bin at any time thanks to the constraint that the total probability of a sample to end in any bin is equal to 1.  
Figure \ref{f:combined-pos} shows the -- average --  medians and 95\% confidence intervals as a function of the number of GW events included in the analysis. The three columns correspond to observations from the HLV network, on the left, from the HLVJ network, on the centre, and from the HLVJI network, on the right. 
Let's focus first on $\Omega_m$ and $\Omega_\Lambda$. Even when we combine information from 50 events, none of these two parameters can be estimated by second generation interferometers, their distance reach is just too small. I did the unrealistic exercise, given the expected rates, of combining all the 1,000 simulated GW events to test whether any kind of information can be extracted and found that not even in that case we can measure the Universe energy densities. We therefore conclude that for a GW-based determination of these two parameters we will have to wait either third generation observatories or space-based ones.
\begin{table*}[!ht]
\begin{tabular}{c|ccc|ccc|ccc}            
\hline\hline
& & HLV & & & HLVJ & & & HLVJI &\\
\hline\hline
\# events & $\langle h_{2.5\%}\rangle$ & $\langle\overline{h}\rangle$ & $\langle h_{97.5\%}\rangle$ & $\langle h_{2.5\%}\rangle$ & $\langle\overline{h}\rangle$ & $\langle h_{97.5\%}\rangle$ & $\langle h_{2.5\%}\rangle$ & $\langle\overline{h}\rangle$ & $\langle h_{97.5\%}\rangle$\\
\hline\hline
5 & $0.644$ & $0.753$  & $0.982$ & $0.664$& $0.701$  & $0.765$ & $0.663$& $0.705$  & $0.779$\\
10 & $0.671$ & $0.714$  & $0.775$ & $0.675$& $0.699$  & $0.725$ & $0.674$& $0.698$  & $0.721$\\
15 & $0.676$& $0.705$  & $0.754$ & $0.681$& $0.699$  & $0.716$ & $0.682$& $0.697$  & $0.712$\\
20 & $0.679$& $0.701$  & $0.722$  & $0.684$& $0.698$ & $0.711$ & $0.684$& $0.697$  & $0.709$\\
30 & $0.681$& $0.698$  & $0.717$ & $0.688$& $0.699$  & $0.708$ & $0.687$ & $0.697$ & $0.707$\\
40 & $0.686$& $0.700$  & $0.714$ & $0.687$& $0.699$  & $0.707$ & $0.689$& $0.697$  & $0.704$\\
50 & $0.686$& $0.700$  & $0.714$ & $0.687$& $0.700$  & $0.706$ & $0.689$& $0.700$  & $0.703$\\
\hline\hline                    
\end{tabular}
\caption{Noise averaged median and 95\% confidence intervals for $h$, averaged over 20 GW event catalogue realisations, as a function of the number of events observed for the three networks under consideration.}
\label{t:h0-events}
\end{table*}

The situation is quite different for $h$ and seemingly for $\Omega_k$. The reduced Hubble constant $h$ will be accurately measured already by second generation instruments with little more than 10 events. Table.~\ref{t:h0-events} reports the average median of $h$ and its 2.5\% and 97.5\% values for the three networks. After as little as 10 events, the measurement of $h$ is accurate to 14.5\%, 7\% and 6.7\% for the HLV, HLVJ and HLVJI networks, respectively. 
The relative uncertainties I find, make GW observations already competitive with results from the Hubble Key Project\cite{FreedmanEtAl:2001} which reports a value for $h$, obtained after combining the results from different methods, of $0.72\pm0.08$ (11\%). With more GW observations the accuracy on $h$ keeps improving, even if not very significantly. However, after 50 GW observations the HLV, HLVJ and HLVJI achieve an accuracy of 5\%, 2\% and 1.8\%, respectively. These measurements have a comparable accuracy to the latest, and most accurate, results available in literature. Komatsu et al. \cite{KomatsuEtAl:2011} combining the information from WMAP, BAO and SnIa, in conjunction with the assumption of a flat universe ($\Omega_k=0$), obtain a best estimate $h = 0.702\pm0.014$ at 1$\sigma$ which, from this study, seems achievable by GW observations alone. 
A closer look at Fig.~\ref{f:combined-pos} reveals that the increase in accuracy cannot go on arbitrarily. The limiting factor is, like for every noise-dominated system, the signal-to-noise ratio. While for the first 10 -- 20 events the uncertainty scales like $\sqrt{N_{\mathrm{obs}}}$, eventually one hits the Cramer-Rao lower bound and no further information can be gained by including further observations.

Other methods relying on the additional assumption that at least one of the components of the coalescing binary system is a neutron star give similar uncertainties. For instance, if one uses only coincident observations of GRBs and GW detections, Nissanke et al.~\cite{NissankeEtAl:2010} find that, with a five interferometers network, $h$ can be measured with $\sim13\%$ fractional error with 4 events, improving to $\sim5\%$ for 15 events. The required number of events increases by $50\%$ and $75\%$ for a four and three instruments network, respectively. Using the neutron star mass function as a statistical mean to extract the redshift of each source, Taylor et al.~\cite{TaylorEtAl:2011} suggest that $h$ can be determined to $\sim10\%$ using $100$ observations. 
%
%
%
%

Quite surprisingly, even if second generation interferometers will not be able to constrain $\Omega_m$ and $\Omega_\Lambda$, Fig.~\ref{f:combined-pos} suggests that, regardless of the network size, we should be able to constrain $\Omega_k$. However, this is most likely not a real effect. In fact, we cannot measure $\Omega_m$ or $\Omega_\Lambda$, therefore their posterior distribution is, just like their prior, a uniform distribution in $[0,1]$. The boundary condition at $z=0$ implies that $\Omega_k = 1 - (\Omega_m + \Omega_\Lambda)$. The distribution of the sum of two uniformly distributed variables in $[0,1]$ is a triangular distribution whose mean is 1. It follows that the mean of $\Omega_k$ if $\Omega_m$ and $\Omega_\Lambda$ are uniformly distributed in [0,1] is 0. The reason for the --apparently-- successful inference of $\Omega_k$ thus relies on the choice of the priors for $\Omega_m$ and $\Omega_\Lambda$. Different choice of the prior ranges would have led to a different inferred value of $\Omega_k$. If we would have chosen uniform priors but within the interval $[0,2]$ instead, we would have inferred for $\Omega_k$ a median value of $-1$. 
Therefore, the convergence of the 95\% confidence intervals towards the correct value $\Omega_k = 0$ is purely a mathematical artifact that follows from the particular choice of the priors. Therefore, we must conclude that second generation interferometers will not constrain $\Omega_k$ either.

\section{Conclusions}\label{s:conclusions} 

This paper presented a scheme for the joint inference of the parameters of a source of GW and any set of cosmological parameters. The scheme herein described allows for a simple inclusion in the analysis of any additional information available. We have seen how previous related studies can be interpreted within this scheme. To exemplify the workings of method, it has been specialised to the study of the expected performance of the upcoming global network of GW observatories. In particular, we compared the two LIGOs and Virgo network with two extended ones including an interferometer in Japan, LCGT, and an interferometer in India, Indigo. In particular, the method has been applied to the case in which the information about the redshift of the sources is obtained using a galaxy catalogue as prior, so for the case in which the redshift is known only statistically. 

Our findings corroborate the fact that adding more instruments to the network substantially increases the accuracy with which the sky parameters can be measured and, most importantly, that the pernicious $D_l$ -- $\iota$ degeneracy can be, at least partially, broken by an extended network. 
However, for the purpose of estimating the cosmological parameters, the three networks behave very similarly. The energy density parameters $\Omega_m$ and $\Omega_\Lambda$ will not be measured by the upcoming network of GW observatories, regardless of their number. In contrast, $h$ will be measured with a precision that is comparable with what is obtained by current electro-magnetic methods. After 10 GW observations, the accuracy (at 95\% confidence) on $h$ is 14.5\%, 7\% and 6.7\% for the HLV, HLVJ and HLVJI networks, respectively, and after 50 GW observations it is 5\%, 2\% and 1.8\%, respectively. Hence, second generation GW detectors will deliver a measurement of the Hubble constant which is comparable with the current value derived from WMAP 7-year observations. 

The very good accuracy that will be obtained for $H_0$ is, in the writer's opinion, only a fortunate collateral effect. The greatest achievement, in the context of cosmology at least, that GW detectors can achieve is an \emph{independent test of the current cosmological paradigm}. Electro-magnetic and GW methods are affected in fact by \emph{entirely different systematics}. The former is afflicted by the curse of the calibration of the distance ladder which relies on a plethora of empirical relations that are either applicable only in a certain range of distances, like the period-luminosity relation for Cepheid stars, or to a certain class of objects, like the Tully-Fisher relation for spiral galaxies or the Faber-Jackson relation for ellipticals. All the various independent methods need to be calibrated against one another to obtain a reliable distance indicator ladder. Further effects also need to be taken into account, like the correction for the galactic interstellar medium absorption and reddening, a similar correction for the observed galaxy interstellar medium and other more subtle effects depending on the details of the actual method.

This problem in GW simply does not exist. The Universe is, for all practical purposes, transparent to GW and, most importantly, the luminosity distance can be observed \emph{directly}. As such, it is not unreasonable to foresee a future in which GW will be the primary calibrator for all other distance indicators, earning not only the name of ``standard sirens'', but also the status.
  
However, GW are not a ``clean'' system. They are still, in fact, affected by some kind of systematics. These include, but are not limited to, uncertainties in the exact shape of the signal, uncertainties in the calibration of the detector, the intrinsic degeneracy between the inclination of the orbital plane of the compact binary and its distance and eventually even uncertainties in the theory of gravity. What the actual consequences on the estimation of the cosmological parameters of the aforementioned sources of bias is not well known or understood. A few studies tried to quantify some of them. According to \cite{VitaleEtAl:2011}, calibration uncertainties induce systematic errors that are typically smaller than the intrinsic statistical uncertainties. However, the study has been performed only for non-spinning systems and whether the conclusion will hold when spins are included in the analysis is still a matter of debate. 
The effect of spins on parameter estimation in general is still uncertain. There are indications that having spinning templates actually improves the accuracy of the inference due to the additional dynamics of the binary system \cite{Vecchio:2004}, but no systematic study is available yet. These potential sources of bias need to be investigated in detail, quantified and, if possible, minimised.

A further possible cause of systematic biases in the particular example of inference of $\cosmoP$ presented in this study, is the incompleteness of the catalogue. When the real host is too faint to be detected by the survey and thus to be considered for the analysis, each single event posterior distribution for sky position, redshift and hence $\cosmoP$, will be displaced compared to the case in which the true host is included. We can get a feeling of the consequences of an incomplete galaxy catalogue by considering two extreme scenarios: (i) the displacement is purely stochastic; (ii) the true host is always further away than what implied by the overall distribution of galaxies in the catalogue. For the former case, the computation of the joint posterior across multiple events will simply average out the single events biases. No special precautions need to be considered and the analysis presented herein remains approximately valid. For the latter case, since the prior on the redshift leads us to always underestimate its value, the joint posterior for $\cosmoP$ will also lead to an underestimate of $H_0$. One might be tempted to consider only events that are louder than some, predetermined, signal-to-noise ratio threshold. Since the signal-to-noise ratio scales essentially like $z^{-2}$, we would be effectively considering sources whose hosts are very unlikely to be missed by our survey. However, the above choice does not yet guarantee that the estimate of $H_0$ would be unbiased. 
The only way to obtain an unbiased estimate of $\cosmoP$ from an incomplete galaxy catalogue is to include in the analysis terms that describe the likelihood of observing a GW whose host was not detected by the survey given its sensitivity. This class of problems -- of which the incompleteness of galaxy catalogues is an example -- is discussed and formally solved in~\cite{MessengerVeitch:2012}. 

Regardless of all the problematic issues raised by the fine details of the measurement process for GW, none of the mentioned potential sources of systematic errors is shared with electro-magnetic methods. A GW-based cosmology is, currently, the only viable way of testing independently what we think we know of the Universe. 

\section*{Acknowledgments} The author is particularly grateful to Alberto Vecchio for the invaluable discussions that led to this study. The author wish to thank T.~G.~F.~Li, C.~Messenger, B.~Sathyaprakash, B.~Schutz, C.~Van Den Broeck, J.~Veitch and S.~Vitale for useful comments and discussions. The numerical simulations were performed on the Tsunami cluster of the University of Birmingham. This research was supported in part by the research programme of the Foundation for Fundamental Research on Matter (FOM), which is partially supported by the Netherlands Organisation for Scientific Research (NWO).

%
%


\begin{thebibliography}{}
\bibitem[Abadie et al.(2010)]{cbc-low-mass-S5VSR1} Abadie, J., Abbott, B.~P., Abbott, R., et al.\ 2010, Classical and Quantum Gravity, 27, 173001 
\bibitem[Aihara et al.(2011)]{SDSSDR8} Aihara, H., Allende Prieto, C., An, D., et al.\ 2011, \apj s, 193, 29 
\bibitem[Althouse et al.(1998)]{AlthouseEtAl:1998} Althouse, W., Jones, L., \& Lazzerini, A., 1998, Technical Report No. LIGO-T980044-08
\bibitem[Anderson et al.(2001)]{AndersonEtAl:2001} Anderson, W.~G., Brady, P.~R., Creighton, J.~D., \& Flanagan, {\'E}.~{\'E}.\ 2001, \prd, 63, 042003 
\bibitem[Belczynski et al.(2008)]{BelczynskiEtAl:2008} Belczynski, K., O'Shaughnessy, R., Kalogera, V., et al.\ 2008, ApJL, 680, L129
\bibitem[Bender(1998)]{lisa} Bender, P.~L.\ 1998, Bulletin of the American Astronomical Society, 30, 1326 
\bibitem[Blanchet et al.(1995)]{BlanchetEtAl:1995} L.~Blanchet, T.~Damour, B.~R.~Iyer, C.~M.~Will and A.~G.~Wiseman, Phys.\ Rev.\ Lett.\  {\bf 74}, 3515 (1995)
\bibitem[Bondu(2010)]{advvirgo} Bondu, F.\ 2010, Gravitation and Fundamental Physics in Space,
\bibitem[Buonanno et al.(2009)]{BuonannoEtAl:2009} Buonanno, A., Iyer, B.~R., Ochsner, E., Pan, Y., \& Sathyaprakash, B.~S.\ 2009, \prd, 80, 084043
\bibitem[Chernoff \& Finn(1993)]{ChernoffFinn:1993} Chernoff, D.~F., \& Finn, L.~S.\ 1993, ApJL, 411, L5
\bibitem[Cutler \& Flanagan(1994)]{CutlerFlanagan:1994}Cutler, C., \& Flanagan, {\'E}.~E.\ 1994, \prd, 49, 2658
\bibitem[Dalal et al.(2006)]{DalalEtAl:2006} Dalal, N., Holz, D.~E., Hughes, S.~A., \& Jain, B.\ 2006, \prd, 74, 063006
\bibitem[Del Pozzo et al.(2011)]{DelPozzoEtAl:2011} Del Pozzo, W., Veitch, J., \& Vecchio, A.\ 2011, \prd, 83, 082002
\bibitem[Deng et al.(2008)]{DengEtAl:2008} Deng X.-F., He J.-Z., Jun S., Luo C.-H., Wu P., 2008, \apj, 51, 471 
\bibitem[Finn(1992)]{Finn:1992} Finn, L.~S.\ 1992, \prd, 46, 5236 
\bibitem[Freedman et al.(2001)]{FreedmanEtAl:2001} Freedman, W.~L., Madore, B.~F., Gibson, B.~K., et al.\ 2001, \apj, 553, 47
\bibitem[Hogg(1999)]{Hogg:1999} Hogg, D.~W.\ 1999, arXiv:astro-ph/9905116 
\bibitem[Holz \& Hughes(2005)]{HolzHughes:2005} Holz, D.~E., \& Hughes, S.~A.\ 2005, \apj, 629, 15
\bibitem[Jackson(2007)]{Jackson:2007} Jackson, N.\ 2007, Living Reviews in Relativity, 10, 4 
\bibitem[Jaynes(2003)]{Jaynes} E.T. Jaynes, Probability Theory: The Logic of Science, Cambridge, Cambridge University Press, 2003
\bibitem[Kissel \& LIGO Scientific Collaboration(2011)]{advligo} Kissel, J.~S., \& LIGO Scientific Collaboration 2011, Bulletin of the American Astronomical Society, \#410.07 
\bibitem[Komatsu et al.(2011)]{KomatsuEtAl:2011} Komatsu, E., Smith, K.~M., Dunkley, J., et al.\ 2011, ApJs, 192, 18
\bibitem[Kuroda \& LCGT Collaboration(2010)]{lcgt} Kuroda, K., \& LCGT Collaboration 2010, Classical and Quantum Gravity, 27, 084004 
\bibitem[MacLeod \& Hogan(2008)]{MacLeodHogan:2008} MacLeod, C.~L., \& Hogan, C.~J.\ 2008, \prd, 77, 043512
\bibitem[Messenger \& Read(2011)]{MessengerRead:2011} Messenger, C., \& Read, J.\ 2011, arXiv:1107.5725
\bibitem[Messenger \& Veitch(2012)]{MessengerVeitch:2012} Messenger, C., \& Veitch, J.\ 2012, arXiv:1206.3461 
\bibitem[Nissanke et al.(2010)]{NissankeEtAl:2010} Nissanke, S., Holz, D.~E., Hughes, S.~A., Dalal, N., \& Sievers, J.~L.\ 2010, \apj, 725, 496
\bibitem[Pannarale et al.(2011)]{PannaraleEtAl:2011} Pannarale, F., Rezzolla, L., Ohme, F., \& Read, J.~S.\ 2011, \prd, 84, 104017 
\bibitem[Petiteau et al.(2011)]{PetiteauEtAl:2011} Petiteau, A., Babak, S., \& Sesana, A.\ 2011, \apj, 732, 82
\bibitem[Sathyaprakash \& Schutz(2009)]{SathyaSchutz:2009} Sathyaprakash, B.~S., \& Schutz, B.~F.\ 2009, Living Reviews in Relativity, 12, 2
\bibitem[Sathyaprakash et al.(2010)]{SathyaEtAl:2010} Sathyaprakash, B.~S., Schutz, B.~F., \& Van Den Broeck, C.\ 2010, Classical and Quantum Gravity, 27, 215006
\bibitem[Sathyaprakash(2011)]{indigo} Sathyaprakash, B.~S. for the LIGO Scientific Collaboration \ 2011 Scientific Benefits of LIGO-India, LSC internal report G1100991
\bibitem[Shoemaker (2009)]{AdvLIGOnoise} D.~Shoemaker (LSC, 2009), \url{https://dcc.ligo.org/cgi-bin/DocDB/ShowDocument?docid=2974} 
\bibitem[Skilling(2004)]{Skilling:2004} Skilling, J.\ 2004, American Institute of Physics Conference Series, 735, 395 
\bibitem[Schutz(1986)]{Schutz:1986} Schutz, B.~F.\ 1986, \nat, 323, 310
\bibitem[Schutz(2011)]{Schutz:2011} Schutz, B.~F.\ 2011, Classical and Quantum Gravity, 28, 125023
\bibitem[Taylor et al.(2011)]{TaylorEtAl:2011} Taylor, S.~R., Gair, J.~R., \& Mandel, I.\ 2011, arXiv:1108.5161 
\bibitem[Vecchio(2004)]{Vecchio:2004} Vecchio, A.\ 2004, \prd, 70, 042001 
\bibitem[Veitch \& Vecchio(2010)]{VeitchVecchio:2010} Veitch, J., \& Vecchio, A.\ 2010, \prd, 81, 062003
\bibitem[Veitch et al.(2012)]{VeitchEtAl:2012} Veitch, J., Mandel, I., Aylott, B., et al.\ 2012, arXiv:1201.1195 
\bibitem[Vitale \& Zanolin(2011)]{VitaleZanolin:2011} Vitale, S., \& Zanolin, M.\ 2011, arXiv:1108.2410 
\bibitem[Vitale et al.(2012)]{VitaleEtAl:2011} Vitale, S., Del Pozzo, W., Li, T.~G.~F. et al.,\ 2011, \prd, 85, 064034 
\bibitem[Wen \& Chen(2010)]{WenChen:2010} Wen, L., \& Chen, Y.\ 2010, \prd, 81, 082001
\bibitem[York et al.(2000)]{YorkEtAl:2000} York D.~G., et al., 2000, AJ, 120, 1579 
\bibitem[Zhao et al.(2011)]{ZhaoEtAl:2011} Zhao, W., van den Broeck, C., Baskaran, D., \& Li, T.~G.~F.\ 2011, \prd, 83, 023005
\end{thebibliography}
\end{document}